\documentclass[
    aps,                
    pra,                
    twocolumn,          
    superscriptaddress, 
    nolongbibliography, 
    nofootinbib,        
    10pt,               
    floatfix            
]{revtex4-2}

\usepackage{graphicx}  
\usepackage{svg}       
\usepackage{float}     
\usepackage{amsmath}   

\usepackage{xcolor}
\usepackage[normalem]{ulem} 
\usepackage{marginnote}     
\usepackage{soul}           

\usepackage{booktabs}  
\usepackage{makecell}  
\usepackage{tikz}      
\usetikzlibrary{matrix,positioning,calc} 
\usepackage{multirow} 
\usepackage{siunitx}   
\DeclareSIUnit{\Gauss}{G}
\DeclareSIUnit{\mrad}{mrad}
\DeclareSIUnit{\mK}{mK}

\usepackage{physics}   
\usepackage{nicefrac}  
\AtBeginDocument{\RenewCommandCopy\qty\SI}

\definecolor{myblue}{HTML}{0b5394}
\usepackage[
    colorlinks=true,
    linkcolor=black,
    citecolor=myblue,
    urlcolor=black
]{hyperref}            

\newcommand{\Ueff}{U_{\text{eff}}}
\newcommand{\figref}[1]{\hyperref[#1]{Fig.~\ref*{#1}}}        
\newcommand{\Eqref}[1]{\hyperref[#1]{equation~(\ref*{#1})}}   
\newcommand{\secref}[1]{\hyperref[#1]{Sec.~\ref*{#1}}}      
\newcommand{\tabref}[1]{\hyperref[#1]{Table~\ref*{#1}}}    

\makeatother
\makeatletter
\renewcommand{\maketitle}{
  \begingroup
    \@author@finish
    \title@column\titleblock@produce
  \endgroup
  \suppressfloats[t]
}
\makeatother

\begin{document}

\newcommand{\MPQ}{Max-Planck-Institut f\"{u}r Quantenoptik, 85748 Garching, Germany}
\newcommand{\MCQST}{Munich Center for Quantum Science and Technology, 80799 Munich, Germany}
\newcommand{\LMU}{Fakult\"{a}t f\"{u}r Physik, Ludwig-Maximilians-Universit\"{a}t, 80799 Munich, Germany}
\newcommand{\Strath}{Department of Physics and SUPA, University of Strathclyde, Glasgow, G4 0NG, United Kingdom}
\newcommand{\LCF}{Laboratoire Charles Fabry, Institut d'Optique Graduate School, CNRS, Universit\'e Paris-Saclay, 91127 Palaiseau, France}

\title{High-fidelity collisional quantum gates with fermionic atoms}
\author{Petar Bojovi\'c}%
\affiliation{\MPQ}%
\affiliation{\MCQST}%
\author{Timon Hilker}%
\affiliation{\MPQ}%
\affiliation{\MCQST}%
\affiliation{\Strath}%
\author{Si Wang}%
\affiliation{\MPQ}%
\affiliation{\MCQST}%
\author{Johannes Obermeyer}%
\affiliation{\MPQ}%
\affiliation{\MCQST}%
\author{Marnix Barendregt}%
\affiliation{\MPQ}%
\affiliation{\MCQST}%
\author{Dorothee Tell}%
\affiliation{\MPQ}%
\affiliation{\MCQST}%
\author{Thomas Chalopin}%
\affiliation{\MPQ}%
\affiliation{\MCQST}%
\affiliation{\LCF}%
\author{Philipp M. Preiss}%
\affiliation{\MPQ}%
\affiliation{\MCQST}%
\author{Immanuel Bloch}%
\affiliation{\MPQ}%
\affiliation{\MCQST}%
\affiliation{\LMU}%
\author{Titus Franz}%
\email[Electronic address: ]{titus.franz@mpq.mpg.de}%
\affiliation{\MPQ}%
\affiliation{\MCQST}%

\begin{abstract}

Quantum simulations of electronic structure and strongly correlated quantum phases are widely regarded as among the most promising applications of quantum computing. 
These computations naturally benefit from native fermionic encodings~\cite{bravyiFermionicQuantumComputation2002, gonzalez-cuadraFermionicQuantumProcessing2023}, which intrinsically restrict the Hilbert space to physical states consistent with fermionic statistics and conservation laws like particle number and magnetization~\cite{gkritsisSimulatingChemistryFermionic2025} independent of gate errors.
While ultracold atoms in optical lattices are established as powerful analog simulators of strongly correlated fermionic matter~\cite{koepsellMicroscopicEvolutionDoped2021, brownBadMetallicTransport2019, hartkeDirectObservationNonlocal2023, xuNeutralatomHubbardQuantum2025}, neutral-atom platforms have concurrently emerged as versatile, scalable architectures for spin-based digital quantum computation~\cite{saffmanQuantumInformationRydberg2010}. Unifying these capabilities requires high-fidelity gates that preserve motional degrees of freedom of fermionic atoms~\cite{murmannTwoFermionsDouble2015c, hartkeQuantumRegisterFermion2022, zhuSplittingConnectingSinglets2025}, paving the way for a new generation of programmable fermionic quantum processors.
Here we demonstrate collisional entangling gates with fidelities up to 99.75(6)\% and Bell state lifetimes exceeding \SI{10}{\s}, realized via controlled interactions of fermionic atoms in an optical superlattice. Using quantum gas microscopy~\cite{chalopinOpticalSuperlatticeEngineering2025}, we microscopically characterize spin-exchange and pair-tunneling gates, and realize a robust, composite pair-exchange gate, a fundamental primitive for quantum chemistry simulations~\cite{mcardleQuantumComputationalChemistry2020, gkritsisSimulatingChemistryFermionic2025}. 
Our results establish controlled collisions in optical lattices as a competitive and complementary approach to high entangling gate fidelities in neutral-atom quantum computers. 
When embedded within a fermionic architecture, this capability enables the preparation of complex quantum states and advanced readout protocols~\cite{impertroLocalReadoutControl2024, schlomerLocalControlMixed2024a, markEfficientlyMeasuring$d$wave2024, tabaresProgrammingOpticallatticeFermiHubbard2025} for a new class of scalable analog-digital hybrid quantum simulators.
Combined with local addressing~\cite{weitenbergSinglespinAddressingAtomic2011, preissStronglyCorrelatedQuantum2015}, these gates mark a crucial step towards a fully digital fermionic quantum computer based on the controlled motion and entanglement of fermionic neutral atoms.

\end{abstract}

\maketitle

Neutral atoms have proven to be a compelling platform for quantum simulation~\cite{blochManybodyPhysicsUltracold2008b, grossQuantumSimulationsUltracold2017, browaeysManybodyPhysicsIndividually2020} and quantum computing~\cite{saffmanQuantumInformationRydberg2010}, offering long lifetimes, inherent scalability, strong tunable interactions, and naturally identical qubits~\cite{saffmanQuantumInformationRydberg2010, henrietQuantumComputingNeutral2020, tabaresProgrammingOpticallatticeFermiHubbard2025}. In the context of spin-based quantum computing, most of the progress to date has been focused on implementations using Rydberg interactions to realize fast and robust quantum gates~\cite{wilkEntanglementTwoIndividual2010, isenhowerDemonstrationNeutralAtom2010, everedHighfidelityParallelEntangling2023, finkelsteinUniversalQuantumOperations2024, munizHighfidelityUniversalGates2024c, radnaevUniversalNeutralatomQuantum2025}.
However, collisional quantum gates were proposed early on as a promising alternative, presenting the potential for high-fidelity operations in both spin and charge degrees of freedom~\cite{jakschEntanglementAtomsCold1999a, brennenQuantumLogicGates1999, sorensenSpinSpinInteractionSpin1999, weitenbergQuantumComputationArchitecture2011a}. 
While first foundational experiments showed the merit of such collisional gates using effective spin-1/2 degrees of freedom encoded in bosonic atoms~\cite{mandelControlledCollisionsMultiparticle2003, follingDirectObservationSecondorder2007, anderliniControlledExchangeInteraction2007a, trotzkyTimeResolvedObservationControl2008}, they lacked single-site and single-spin sensitive readout capabilities and were thereby limited in the microscopic assessment of realized gate fidelities. 
With the advent of quantum gas microscopy, a better characterization of such collisional quantum gates became possible, including the realization of complex multi-particle entangled quantum states~\cite{zhangScalableMultipartiteEntanglement2023b}.

The highest reported two-particle entangling gate fidelities from such microscopic analysis reached approximately $F \sim 96\% $~\cite{zhangScalableMultipartiteEntanglement2023b},
whereas earlier experiments without microscopic resolution reported fidelities as high as $F \sim 99.3\% $~\cite{yangCoolingEntanglingUltracold2020}.

\begin{figure}[!t]
	\centering
	 \includegraphics[scale=0.9751]{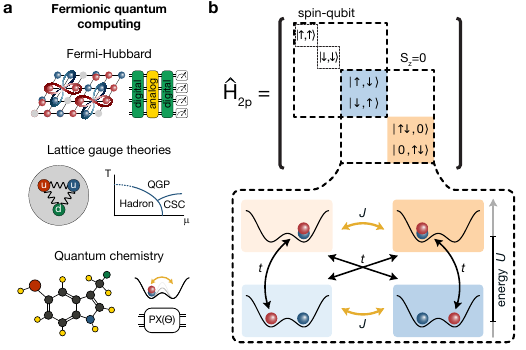}
	\caption{
		\textbf{Fermionic quantum processor}:
		\textbf{a,} 
		By controlling the spin and motional dynamics of fermionic atoms with digital gates, a future lattice-based quantum processor can efficiently simulate strongly correlated systems of many electrons or other fermionic particles.
		\textbf{b,} At its core, double-well potentials of an optical superlattice are used to entangle particles.
		In the two-particle sector of the Fermi-Hubbard Hamiltonian, the single-particle tunneling processes $t$, detuned by the on-site interaction $U$, give rise to the spin-exchange and pair-tunneling, both characterized by an effective coupling $J$.
		Through appropriate experimental control sequences, the resulting dynamics realize high-fidelity $\mathrm{SWAP}^\alpha$ gates.
	}         
	\label{fig:fig1}
\end{figure}

In this work, we demonstrate entangling collisional gates using fermionic $^6\mathrm{Li}$ atoms in both spin and charge degrees of freedom---essential building blocks for quantum computing architectures based on fermions.
Using quantum gas microscopy, we determine both the continuous-time and discrete gate-based performance of the collisional interactions. 
For the latter, we find high entangling gate fidelities up to $99.75(6)\%$.
Additionally, we generate entangled Bell states and observe noise-resilient entanglement with lifetimes exceeding 10 seconds.
We find that the primary source of gate infidelity originates from averaging over double-wells with slightly different oscillation frequencies, indicating strong prospects for achieving even higher fidelities using, e.g., echo pulses, optical potential flattening procedures~\cite{sompetRealizingSymmetryprotectedHaldane2022, hirtheMagneticallyMediatedHole2023a, wangHomogeneousFermionicHubbard2025}
and optimal control sequences~\cite{nemirovskyFastUniversalTwoqubitGate2021, singhOptimizingTwoqubitGates2025}.
Furthermore, we engineered a pulse sequence that decouples spin-exchange from pair-tunneling dynamics, thereby enabling a pair-exchange gate.
In conjunction with single-particle hopping~\cite{chalopinOpticalSuperlatticeEngineering2025, impertroLocalReadoutControl2024}, this mechanism is an essential building block for future digital fermionic quantum computers with applications in materials science and chemistry~\cite{gonzalez-cuadraFermionicQuantumProcessing2023,  gkritsisSimulatingChemistryFermionic2025, anselmettiLocalExpressiveQuantumnumberpreserving2021, yoshiokaHuntingQuantumclassicalCrossover2024} (\figref{fig:fig1}a bottom).
Such a future device encodes fermionic problems more directly thus naturally restricting the Hilbert space to the one for fermionic particles, thereby intrinsically preventing any non-physical states.
In addition, it conserves particle number as well as polarization independent of gate errors. 
In the near term, these gates already enable novel readout and preparation schemes of strongly correlated electronic states~\cite{schlomerLocalControlMixed2024a, markEfficientlyMeasuring$d$wave2024, tabaresProgrammingOpticallatticeFermiHubbard2025} in hybrid analog-digital quantum simulation schemes (\figref{fig:fig1}a top). 
In the long term, fermion-based architectures also hold the potential to simulate the dynamics of lattice gauge theories~\cite{gonzalez-cuadraFermionicQuantumProcessing2023, zacheFermionquditQuantumProcessors2023} (\figref{fig:fig1}a middle).

The dynamics of spin-1/2 fermions confined in a double-well potential is  captured by the two-site ($\mathrm{L}$, $\mathrm{R}$) Fermi-Hubbard Hamiltonian
\begin{align}
    \hat{H}_{\mathrm{FH}} = &-t\sum_{\sigma} \left(\hat{c}_{\mathrm{L}, \sigma}^{\dagger} \hat{c}_{\mathrm{R}, \sigma} + \mathrm{h.c.}\right) + U \sum_{i}\hat{n}_{i, \uparrow}\hat{n}_{i, \downarrow} \notag \\
    &+ \frac{\delta}{2} \sum_{\sigma} (\hat{n}_{\mathrm{R}, \sigma}-\hat{n}_{\mathrm{L}, \sigma})  + \Delta_B(\hat{n}_{\mathrm{R}, \uparrow} - \hat{n}_{\mathrm{R}, \downarrow})
\label{eq:H_FH}    
\end{align}
where $t$ is the particle tunneling energy, $U$ the on-site repulsive interaction, and $\delta$ indicates a spin-independent potential offset between left and right sites of the double-well. 
A spin-dependent chemical potential gradient, induced by an applied magnetic field gradient, is denoted by $\Delta_B$. 
The operator $\hat{c}_{i,\sigma}$ ($\hat{c}_{i,\sigma}^\dagger$) annihilates (creates) a fermion with spin $\sigma \in \{\uparrow, \downarrow\}$ on site $i \in \{\mathrm{L}, \mathrm{R}\}$, and $\hat{n}_{i,\sigma}$ represents the corresponding density operator.
In the minimal configuration relevant for collisional gates---two fermions in the double-well---the Hamiltonian $\hat{H}_{\mathrm{FH}}$ effectively reduces to a six-dimensional Hilbert space. 
Within this space, $\hat{H}_{\mathrm{FH}}$ acts non-trivially only on the four-dimensional subspace with total spin projection $S_z = 0$, where spin-exchange and correlated pair-tunneling naturally emerge as second-order processes with identical exchange amplitude $J$ (\figref{fig:fig1}b, for the general expression see the Supplementary Information (SI)).

Full control over the motional degrees of freedom of two fermions in a double-well is a key prerequisite for digital fermionic quantum computation, requiring the ability to isolate individual tunneling from exchange processes.
This is achieved experimentally by implementing a unitary interaction matrix in the basis $\{\ket{\uparrow, \uparrow}, \ket{\downarrow, \downarrow}, \ket{\uparrow, \downarrow}, \ket{\downarrow,\uparrow}, \ket{\uparrow\downarrow, 0}, \ket{0, \uparrow\downarrow}\}$:
\begin{equation}
    U_{\mathrm{int}}(\theta) =
    \left[\begin{smallmatrix}
        1 & 0 & 0 & 0 & 0 & 0\\ 
        0 & 1 & 0 & 0 & 0 & 0\\
        0 & 0 & \frac{1 + e^{i\theta}}{2} & \frac{1 - e^{i\theta}}{2} & 0 & 0\\
        0 & 0 & \frac{1 - e^{i\theta}}{2} & \frac{1 + e^{i\theta}}{2} & 0 & 0\\
        0 & 0 & 0 & 0 & e^{-i\zeta}\frac{1 + e^{-i\theta}}{2} & -e^{-i\zeta}\frac{1 - e^{-i\theta}}{2}\\
        0 & 0 & 0 & 0 & -e^{-i\zeta}\frac{1 - e^{-i\theta}}{2} & e^{-i\zeta}\frac{1 + e^{-i\theta}}{2}
    \end{smallmatrix}\right],
    \label{eq:XXZZgate}
\end{equation}
where the angle $\theta$ is determined by the exchange coupling $J$ and the gate duration $\tau_\mathrm{h}$, while the angle $\zeta$ depends on both the on-site interaction $U$ and $\tau_\mathrm{h}$. For a quench, $\theta = 2 \pi \times J \tau_\mathrm{h} / h$ and $\zeta = 2 \pi \times U \tau_\mathrm{h} / h$, where $h$ is Planck's constant (for a detailed derivation see SI).

\begin{figure}[!t]
	\centering
	 \includegraphics[scale=0.9751]{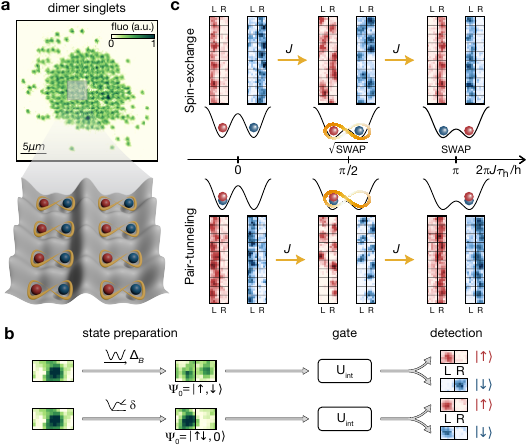}
	\caption{
		\textbf{Two-particle gates in double-wells}:
		\textbf{a,} 
		Example shot of dimer singlets in a double-well lattice
		\textbf{b,} 
		Experimental sequence for N\'eel or doublon initial state, 
		which are prepared by applying a spin-dependent chemical potential $\Delta_B$ (top) or a double-well tilt $\delta$ (bottom).
		\textbf{c,} Experimental shots of a $2\times 10$ subsystem showing continuous $\textrm{SWAP}^\alpha$ evolution from the initial product state (left) through the entangled state (middle) to the swapped product state (right) in the spin (top) and doublon (bottom) sector.
	}
	\label{fig:fig2}
\end{figure}

\begin{figure*}[!t]
	\centering
	\includegraphics[scale=0.9754]{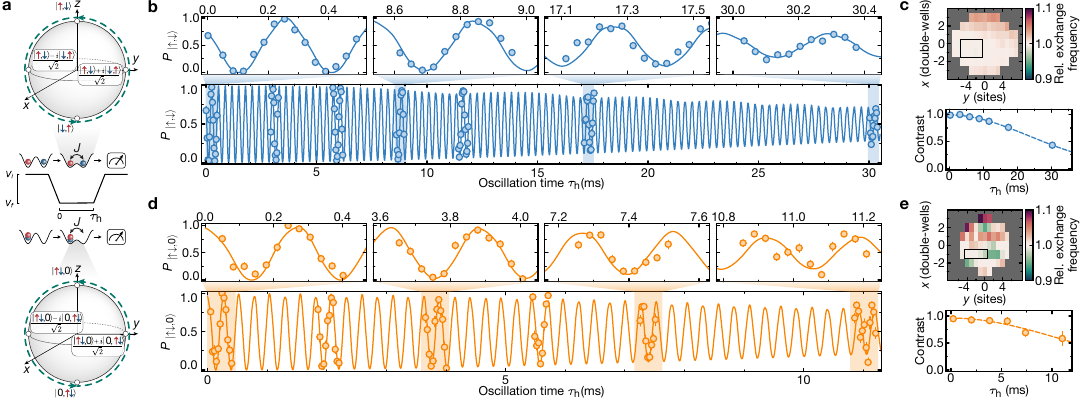}
	\caption{
		\textbf{Coherent spin-exchange and pair-tunneling}:
		\textbf{a,} 
		Evolution of the initial state $\ket{\uparrow, \downarrow}$ (top) and $\ket{\uparrow\downarrow,0}$ (bottom) on a two-particle Bloch sphere under exchange $J$ controlled by the lattice depths (middle).
		\textbf{b, d,} 
		The populations of the states $\ket{\uparrow, \downarrow}$ and $\ket{\uparrow\downarrow, 0}$, each evolving under exchange interactions with frequencies $J/h = \SI{3.32(3)}{\kilo\hertz}$ and $J/h = \SI{3.8(2)}{\kilo\hertz}$, respectively.
		The outsets show enlarged sections of the full data set.
		The data is post-selected on double-wells that only contain $\ket{\uparrow, \downarrow}$ and $\ket{\downarrow, \uparrow}$ (in \textbf{b}), and $\ket{\uparrow\downarrow, 0}$ and $\ket{0, \uparrow\downarrow}$ (in \textbf{d}).
		We find good agreement with numerical simulations of the Fermi-Hubbard Hamiltonian \Eqref{eq:H_FH} (solid lines), which exhibit a small anharmonicity due to a finite ramp speed.
		\textbf{c,} 
		Top: Map of relative exchange oscillation frequency per double-well with a black rectangle indicating the region of the system used in~\textbf{b}. 
		Bottom: The decay of contrast in the region of interest (data points) is well described by a fitted Gaussian-type decay time of $\tau_{\mathrm{ex}}=\SI{33(2)}{ms}$ (dashed line).
		\textbf{e,} 
		same as in \textbf{c}, but for $\ket{\uparrow\downarrow,0}$ state.
		The fitted Gaussian-type decay is $\tau_{\mathrm{ex}}=\SI{15(3)}{ms}$.
		All error bars are calculated as a 68\% confidence interval, unless stated otherwise.
	}
	\label{fig:fig3}
\end{figure*}

The upper-diagonal 4×4 block captures the unitary dynamics of the spin degree of freedom, effectively realizing a $\mathrm{SWAP}^\alpha$ ($\alpha=\theta/\pi$) gate within this subspace. 
For $\alpha = 1/2$, this implements the entangling $\sqrt{\mathrm{SWAP}}$ gate, which, together with single-qubit gates, forms a universal gate set for spin-based quantum computers \cite{hayesQuantumLogicExchange2007, nielsenQuantumComputationQuantum2012}.
The lower-diagonal 2×2 matrix describes coherent pair-tunneling, which entangles two particles in their charge degrees of freedom.

In our experiment, we load a Fermi-degenerate gas of $^6\mathrm{Li}$ atoms into a two-dimensional square optical lattice to prepare a state with two atoms of opposite spins at most lattice sites~\cite{xuNeutralatomHubbardQuantum2025}. 
Individual lattice sites are split into double-wells by superimposing a lattice in the \textit{x}-direction with exactly half the lattice spacing (short lattice), forming an optical superlattice. 
Lattice depths are expressed in units of their respective recoil energy $E_r = h^2/8ma^2$, with atomic mass $m$ and respective lattice constant $a$.
By setting the double-well bias $\delta = 0$, a symmetric double-well potential is formed, resulting in the splitting of doublons into dimer singlets (\figref{fig:fig2}a) \cite{chalopinOpticalSuperlatticeEngineering2025, zhuSplittingConnectingSinglets2025}.
Similarly, we can prepare the highlighted eigenstates in \figref{fig:fig1}b: To initialize a product state $\ket{\Psi_0} = \ket{\uparrow, \downarrow}$, we apply a strong magnetic field gradient along the \textit{x}-direction during splitting while maintaining the symmetric configuration (\figref{fig:fig2}b, upper plot)~\cite{bollSpinDensityresolvedMicroscopy2016a}.
The states $\ket{\uparrow\downarrow, 0}$ and $\ket{0, \uparrow\downarrow}$ are initialized by introducing a controlled offset $\delta$ while ramping up the short lattice (\figref{fig:fig2}b, lower plot)~\cite{follingDirectObservationSecondorder2007}. 
After a quantum gate is applied (\figref{fig:fig2}c), we perform a projective measurement and detect the final state using spin and charge site-resolved fluorescence imaging~\cite{koepsellRobustBilayerCharge2020a}.

We first probe continuous spin-exchange and pair-tunneling dynamics. 
The spin-exchange interaction, which couples the states $\ket{\uparrow, \downarrow}$ and $\ket{\downarrow, \uparrow}$, is initiated by lowering the intra-double-well barrier in \SI{500}{\micro\second}.
The resulting dynamics are visualized on a two-particle Bloch sphere (\figref{fig:fig3}a, top) as a rotation around the \textit{x}-axis where the entangled state $\tfrac{1}{\sqrt{2}}(\ket{\uparrow, \downarrow} - i\ket{\downarrow, \uparrow})$ is populated after a $\theta=\pi/2$ rotation. 
For our parameters (see SI), we observe long-lived oscillations of the population $\ket{\uparrow, \downarrow}$ with a frequency of $J/h = \SI{3.32(3)}{\kilo\hertz}$ (\figref{fig:fig3}b) on twenty central lattice sites.
From the Gaussian decay of the contrast of these oscillations $\propto e^{-(\tau_\mathrm{h}/\tau_{\mathrm{ex}})^2}$ (\figref{fig:fig3}c, bottom), we extract a $1/e$ decay time of $\tau_{\mathrm{ex}} = \SI{33(2)}{\milli\second}$, corresponding to a quality factor of $J\tau_{\mathrm{ex}}/h = 110(8)$ coherent oscillations---the highest recorded in superlattice platforms~\cite{yangCoolingEntanglingUltracold2020} and for any collisional entangling gates~\cite{murmannTwoFermionsDouble2015c, kaufmanEntanglingTwoTransportable2015}.
A related two-particle exchange dynamics can be realized in the charge degree of freedom when the system is initialized in the state $\ket{\uparrow\downarrow,0}$. 
This state is coupled to $\ket{0, \uparrow\downarrow}$ (see Bloch sphere in \figref{fig:fig3}a, bottom), giving rise to coherent pair-tunneling~\cite{winklerRepulsivelyBoundAtom2006, follingDirectObservationSecondorder2007}.
Constraining the analysis to a region of ten lattice sites and post-selecting on measurements with only these two states, we extract a $1/e$ decay time of $\tau_{\mathrm{ex}} = \SI{15(3)}{\milli\second}$, corresponding to a record-high $J\tau_{\mathrm{ex}}/h = 55(10)$ coherent oscillations in the population of $\ket{\uparrow\downarrow, 0}$ (\figref{fig:fig3}d,e).

To assess the effect of spatial inhomogeneity on the observed coherences, we extract maps of the local gate oscillation frequency (\figref{fig:fig3}c and \figref{fig:fig3}e, upper panels). 
We find that the spatial variation in frequency is consistent with the observed decoherence time, indicating that in both cases spatial inhomogeneity is the dominant limitation of the quality factor (see SI).
The essential distinction between spin-exchange and pair-exchange lies in the increased sensitivity of the latter to motional dynamics: Any tilt $\delta$ in the double-well potential, whether caused by phase shifts between the short and long lattices or by lattice beam inhomogeneities, lifts the degeneracy between the states $\ket{\uparrow\downarrow,0}$ and $\ket{0,\uparrow\downarrow}$. This results in dephasing that scales as $(\delta/J)^2$, in contrast to the $(\delta/U)^2$ scaling characteristic of spin-exchange.
The heightened sensitivity accounts for both the pronounced spatial dependence of the pair-exchange frequencies and the more rapid decay of contrast (\figref{fig:fig3}c and \figref{fig:fig3}e, lower panels).

Coherent spin-exchange interactions in a double-well potential can be harnessed to realize two-qubit $\mathrm{SWAP}^\alpha$ gates, as shown in \Eqref{eq:XXZZgate}, with the relevant Hilbert space formed by $\{\ket{\uparrow, \uparrow}, \ket{\downarrow, \downarrow}, \ket{\uparrow, \downarrow}, \ket{\downarrow, \uparrow}\}$.
To implement this in the experiment, we need to suppress single-particle tunneling events that mix the spin sector \{$\ket{\uparrow, \downarrow}, \ket{\downarrow,\uparrow}$\} with charge sector \{$\ket{\uparrow\downarrow, 0}, \ket{0, \uparrow\downarrow}$\}. 
This can be achieved in the regime $U/t \gg 1$ where such processes become far off-resonant at the cost of reducing gate speed $J\approx4t^2/U$. 
An approach leading to faster gate operation is to fine-tune $U/t$ to a magic ratio $4/\sqrt{3}$~\cite{yangCoolingEntanglingUltracold2020} (see Supplementary Information).
For this ratio, the effective single-particle tunneling rate is exactly four times larger than $J$, leading to decoupling of the spin and charge sectors at integer multiples of  $\tau_\mathrm{h} = h/(4J)$.
This is confirmed in our measurements (\figref{fig:fig4}a left), where we quench the dynamics by ramping down the short lattice depth over \SI{50}{\micro\second}, a duration that is fast compared to the inverse of the energy gap $U$, thereby reaching a regime with $U/t = 4/\sqrt{3}$.
We observe a significant fraction of up to 40\% of doublons excited during the time evolution $\tau_\mathrm{h}$, which drops to less than 5\% for a $\theta = \pi/2$ entangling $\sqrt{\mathrm{SWAP}}$ pulse (dashed lines) or $\theta=\pi$ spin-exchange $\mathrm{SWAP}$ pulse (dotted lines)~\cite{yangCoolingEntanglingUltracold2020}.
However, the required fine-tuning of the gate parameters can make it less robust to technical imperfections.

In an alternative approach, we introduce an intermediate-speed strategy in which the tunneling $t$ is ramped slowly compared to $U$ but still fast with respect to $J$. 
This avoids mixing the spin and charge sectors without the need to keep $U/t \gg 1$, or to fine-tune $U/t$ (see Supplementary Information).
In this case, the total doublon fraction never exceeds \SI{8}{\percent}, independent of gate duration (\figref{fig:fig4}a bottom middle).
Stronger suppression of doublon excitations to lower than \SI{5}{\percent} can be achieved by using smoothly shaped Blackman pulses longer than \SI{0.75}{\milli\second}
~(\figref{fig:fig4}a, right). 
We ramp the short lattice depth from $54\,E_r^{\rm short}$ to $6.50\,E_r^{\rm short}$, reaching a minimum of $U/t \simeq 2.85$, with the long lattice depth held constant at $38.0\,E_r^{\rm long}$.
As a result of the high intrinsic energy scales in our system, our $\sqrt{\mathrm{SWAP}}$ gate duration of \SI{1.2}{\ms} (dashed line) remains notably faster than previously reported durations~\cite{mandelControlledCollisionsMultiparticle2003, follingDirectObservationSecondorder2007, anderliniControlledExchangeInteraction2007a, trotzkyTimeResolvedObservationControl2008, yangCoolingEntanglingUltracold2020}.
This quasi-adiabatic approach based on Blackman pulses provides a good compromise between robustness and speed. Hence, we employ it in the remainder of this section.
\figref{fig:fig4}b shows the truth table for a single $\sqrt{\mathrm{SWAP}}$ pulse applied to $\ket{\uparrow,\downarrow}$ and $\ket{\downarrow,\uparrow}$.
The spin-polarized states $\ket{\uparrow, \uparrow}, \ket{\downarrow, \downarrow}$ are not coupled to any other state by the Hamiltonian of \Eqref{eq:H_FH}, therefore, the theoretically expected final occupations for these initial states are shown in gray. 

\begin{figure}[!t]
	\centering
	 \includegraphics[scale=0.9751]{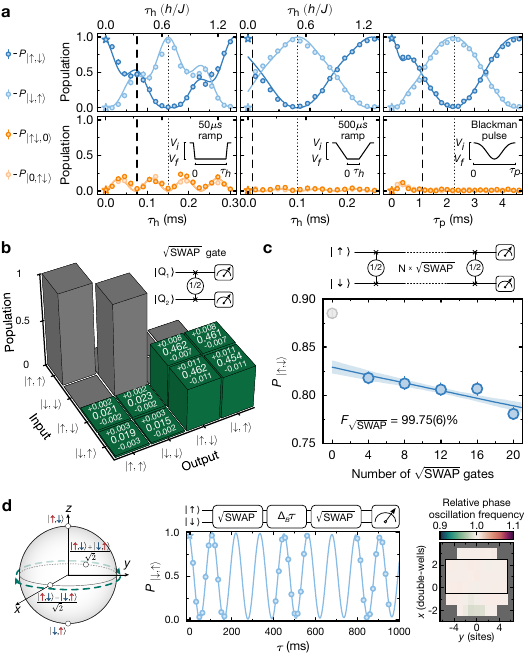}
	\caption{
		\textbf{High-fidelity $\mathrm{SWAP}^\alpha$ gates}: 
		\textbf{a,} Populations of states \{$\ket{\uparrow,\downarrow}, \ket{\downarrow,\uparrow}, \ket{\uparrow\downarrow,0}, \ket{0,\uparrow\downarrow}\}$ in a double-well as a function of pulse duration for three different pulse ramp shapes: \SI{50}{\micro\second} linear (left), \SI{500}{\micro\second} linear (center) and Blackman pulse (right) (see insets). 
		In all cases, the system is initialized in $\ket{\uparrow, \downarrow}$ (indicated by star symbols).
		For linear pulses, we keep the ramp duration fixed and vary the hold time $\tau_h$, while for the Blackman pulse, we scan the total pulse duration $\tau_p$. 
		The pulse durations corresponding to $\sqrt{\text{SWAP}}$ ($\text{SWAP}$) gate are indicated by dashed (dotted) vertical lines. 
		The solid lines are obtained from the simulation.
		A secondary \textit{x}-axis shows the pulse duration in units of maximal spin-exchange frequency $h/J$.
		\textbf{b,} Measured truth table for $\sqrt{\mathrm{SWAP}}$ gate for input states $\ket{\uparrow,\downarrow}$ and $\ket{\downarrow,\uparrow}$. 
		Input states $\ket{\uparrow,\uparrow}$ and $\ket{\downarrow,\downarrow}$ are not coupled by the natural Hamiltonian, and the theory expectations are shown in gray.
		\textbf{c,} Return probability after multiple $\sqrt{\mathrm{SWAP}}$ gates.
		The two-qubit gate fidelity of $99.75(6)\%$ is extracted from an exponential decay fit of the state population (solid blue line). The uncertainty on fidelity is estimated using bootstrapping.
		\textbf{d.} Singlet-triplet oscillations are driven by a magnetic field gradient along the double-well potential axis. 
		The data is well fitted by a damped sinusoidal with a lower bound on coherence time of $\SI{10}{s}$, well beyond the measurement duration (see SI). 
		A spatial map of measured Ramsey frequencies is shown on the right.
		All truth tables are shown without SPAM correction.
	}
	\label{fig:fig4}
\end{figure}

To extract the two-qubit gate fidelity, we apply up to 20 consecutive gates and fit an exponential decay to the population of the $\ket{\uparrow, \downarrow}$ state (\figref{fig:fig4}c). 
We analyze the data on 64 lattice sites, with over 175 experimental realizations per setting, and post-select on double-wells occupied by two particles.
The offset extracted from the fit quantifies the state preparation and measurement (SPAM) error. 
The grayed-out data point, representing the measurement without any applied gate, shows artificially increased population as band excitations from the state preparation are initially misidentified and relax after the first applied gate (see SI). 
These SPAM errors also impact the truth table measurement. 
The fitting function accurately captures the remaining data points, yielding an average gate fidelity of 99.75(6)\% for entangling two adjacent neutral atoms in the entire array. 
Using more sophisticated randomized benchmarking protocols, similar fidelities have been reported in fermionic platforms based on quantum dots~\cite{noiriFastUniversalQuantum2022a, xueQuantumLogicSpin2022}, for individual double quantum dot devices.

To evaluate the lifetime of the entangled state, we perform a spin-sensitive Ramsey measurement. 
Starting from the $\ket{\uparrow, \downarrow}$ state, we apply a $\sqrt{\mathrm{SWAP}}$ gate to entangle the atoms. 
A strong magnetic field gradient of $\SI{40}{\Gauss\cm^{-1}}$ is then applied along the double-well potential, at a Fesh\-bach field of \SI{686.9}{\Gauss}.
This lifts the degeneracy between the $\ket{\uparrow, \downarrow}$ and $\ket{\downarrow, \uparrow}$ states by \SI{9}{\Hz}, resulting in a phase rotation of the entangled state.
On the two-particle Bloch sphere, this corresponds to a rotation around the \textit{z}-axis (\figref{fig:fig4}d left), inducing singlet-triplet oscillations~\cite{greifShortRangeQuantumMagnetism2013}.
A second disentangling $\sqrt{\mathrm{SWAP}}$ pulse subsequently enables state readout in the single-particle basis $\{ \ket{\uparrow}, \ket{\downarrow} \}$ (\figref{fig:fig4}d middle).
The oscillations show close to full contrast over one second of measurement time, with thermal effects in the experiment limiting longer durations.
We estimate (see SI) a coherence time of more than \SI{10}{\second}, which is four orders of magnitude longer than the duration of the single entangling gate.
This long coherence time is ensured by the small differential magnetic dipole moment in the Paschen-Back regime.

Much of the untapped potential of fermions in optical superlattices lies outside the demonstrated and robust spin-qubit framework.
Notably, controlling fermionic motional degrees of freedom via gates enables opportunities for the use of ultracold fermions in material science and quantum chemistry.
In this context, a common approach is to endow an uncorrelated Hartree-Fock wavefunction with correlations via coherent excitations \cite{mcardleQuantumComputationalChemistry2020, gkritsisSimulatingChemistryFermionic2025}. 
These can be single excitations, driven by single-particle tunneling, or double excitations, created by correlated particle motion.

Using concatenated gate sequences, we demonstrate the engineering of specific double excitations. We focus on the pair-exchange process $\mathrm{PX}(\Theta)$, which describes the correlated tunneling of an unbroken fermionic pair between two spatial orbitals,

\begin{equation}
    \mathrm{PX}(\Theta) =
    {
    \begin{tikzpicture}[baseline=(M.center)]
        \matrix (M) [%
          matrix of math nodes,
          left delimiter={[}, right delimiter={]},
          nodes in empty cells,
          nodes={%
            font=\small,              
            minimum width=0.6em,      
            minimum height=0.6em,     
            anchor=center
          },
          column sep=0.2em,           
          row sep=0.2em               
          ]{
             1 & 0 & 0 & 0 & \phantom{0} & \phantom{0} \\
             0 & 1 & 0 & 0 & \phantom{0} & \phantom{0} \\
             0 & 0 & 1 & 0 & \phantom{0} & \phantom{0} \\
             0 & 0 & 0 & 1 & \phantom{0} & \phantom{0} \\
             \phantom{0} & \phantom{0} & \phantom{0} & \phantom{0}
               & \cos(\Theta)
               & -\sin(\Theta) \\
             \phantom{0} & \phantom{0} & \phantom{0} & \phantom{0}
               & \sin(\Theta)
               & \cos(\Theta) \\
          };
          \node at ($(M-1-5)!0.5!(M-4-6)$) {\large$\mathbf{0}$};
          \node at ($(M-5-1)!0.5!(M-6-4)$) {\large$\mathbf{0}$};
    \end{tikzpicture}
    }\,.
    \label{eq:PXgate}
\end{equation}

\begin{figure}[]
	\centering
	 \includegraphics[scale=0.9751]{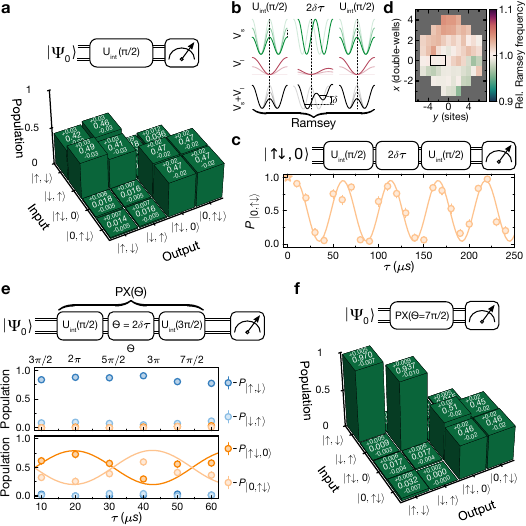}
	\caption{
		\textbf{Pair-exchange composite gate sequence}:
		\textbf{a,} Experimental truth table for the lower-diagonal 4×4 block of the interaction matrix $U_{\mathrm{int}}$ with $\theta=\pi/2$
		\textbf{b,} Superlattice potentials (short lattice in green, long lattice in red, and total potential in black) used to implement the charge-sensitive Ramsey sequence composed of two $U_\mathrm{int}$ and one $Z$-gate tilt.
		\textbf{c,} 
		Ramsey oscillation in the population of $\ket{\uparrow\downarrow, 0}$ for the sequence in \textbf{b}. 
		\textbf{d,}	Map of relative Ramsey frequencies per double-well with a black rectangle indicating the region of the system used in \textbf{c}, with an extracted frequency of \SI{19.2(2)}{\kilo \hertz}. 
		\textbf{e,}	Applying a circuit of interaction- and tilt-pulses for the initial states $\ket{\Psi_0} = \ket{\uparrow, \downarrow}$ (top) and $\ket{\Psi_0} = \ket{\uparrow\downarrow, 0}$ (bottom). This generates an effective pulse that only performs a coherent pair-tunneling $\ket{\uparrow\downarrow, 0} \leftrightarrow \ket{0, \uparrow\downarrow}$ without changing the spin states $\ket{\uparrow, \downarrow}$ and $\ket{\downarrow, \uparrow}$.
		The solid lines in \textbf{c,} and \textbf{e,} correspond to the simulated data.
		\textbf{f,} 
		Experimental truth table of the pair-exchange gate $\mathrm{PX}(\Theta=7\pi/2)$.}
	\label{fig:fig5}
\end{figure}

While the Fermi-Hubbard model naturally realizes coherent pair-tunneling~\cite{follingDirectObservationSecondorder2007} as demonstrated by the oscillations in \figref{fig:fig3}d, this process is usually coupled to other dynamics such as spin-exchange or single-particle tunneling. 
We develop and demonstrate a composite gate that specifically isolates the pair-exchange gate from these other processes (see SI). 
The main component of the composite gate consists of interaction dynamics that generate a discrete gate within the lower-diagonal 4×4 block of the unitary operator in \Eqref{eq:XXZZgate}. 
The truth table of a quasi-adiabatic Blackman interaction gate with rotation angle $\theta = \pi/2$ is shown in \figref{fig:fig5}a.
We then combine the interaction gate with a charge-sensitive dynamical tilt $\delta$ of the double well ($Z$-gate, see \Eqref{eq:H_FH}).
To reduce the sensitivity to spatial inhomogeneities and temporal fluctuations, we devise a new experimental protocol for these gate sequences (see \figref{fig:fig5}b and SI). 
Since phase errors scale with the depth of the long lattice, we generate the potential tilt $\delta$ near the maximally staggered configuration, where even a shallow long lattice produces a strong differential tilt, making small phase errors significantly less impactful compared to previous experimental approaches~\cite{impertroLocalReadoutControl2024}. A deep long lattice potential is used only during the $U_\mathrm{int}$ pulses to confine atoms within each double well.
The optimized $Z$-gate enables the observation of fast and coherent Ramsey oscillations (\figref{fig:fig5}c) with a frequency of tens of \SI{}{\kilo \hertz}. 
Residual spatial variation in Ramsey frequencies across the system (\figref{fig:fig5}d) can be used to map out the spatial inhomogeneity of $\delta$.

We implement the pair-exchange gate $\mathrm{PX}(\Theta)$ by interleaving $\theta = \pi/2$ and $\theta = 3\pi/2$ interaction pulses with a $Z$-gate. 
The states $\ket{\uparrow, \downarrow}$, $\ket{\downarrow, \uparrow}$ (light and dark blue circles in \figref{fig:fig5}e top) are not affected by the tilt $\delta$ and perform a full $\theta=2\pi$ rotation such that the population remains constant. 
On the other hand, for initial states $\ket{\uparrow\downarrow, 0}$ and $\ket{0, \uparrow\downarrow}$, the final state depends on the double-well tilt $\delta$, and we observe an oscillation with a frequency of $2 \delta$ (\figref{fig:fig5}e bottom). 
The reduced contrast of doublon oscillations originates from direct spin-exchange in the extended Fermi-Hubbard model~\cite{trotzkyTimeResolvedObservationControl2008, follingDirectObservationSecondorder2007}, that slightly modifies the interaction gate in~\Eqref{eq:XXZZgate}.
We illustrate the effect of the sequence by the truth table of a $\mathrm{PX}(\Theta=2\delta\tau=7\pi/2)$ gate shown in \figref{fig:fig5}f. 
Unlike the native interaction gate (\figref{fig:fig5}a), this gate selectively performs a $\sqrt{\mathrm{SWAP}}$ on the charge sector while leaving the spin sector unaffected. 
This proof-of-principle realization marks an important step towards utilizing fermions in optical superlattices for fermionic quantum computing, for instance for the simulation of non-native Hamiltonians~\cite{tabaresProgrammingOpticallatticeFermiHubbard2025} or variational methods in quantum chemistry~\cite{gkritsisSimulatingChemistryFermionic2025}.

The high-fidelity collisional entangling gates and the intrinsically long coherence times demonstrated here establish optical lattices as a compelling route toward scalable quantum computing. 
Reconfiguring the superlattice dimerization extends the platform beyond isolated dimers, enabling, for example, the generation of large-scale entanglement~\cite{zhangScalableMultipartiteEntanglement2023b}.
Collisional gate performance can be further improved through optical potential flattening \cite{sompetRealizingSymmetryprotectedHaldane2022, hirtheMagneticallyMediatedHole2023a, wangHomogeneousFermionicHubbard2025} and optimal control techniques \cite{everedHighfidelityParallelEntangling2023, singhOptimizingTwoqubitGates2025}. 
One can readily anticipate sub-\SI{10}{\micro\second} entangling gates acting on systems as large as 10,000 lattice sites (see SI). 
Single-qubit gates and local control, as already realized in advanced Rydberg-based neutral-atom platforms \cite{saffmanQuantumComputingNeutral2019, everedHighfidelityParallelEntangling2023, munizHighfidelityUniversalGates2024c, finkelsteinUniversalQuantumOperations2024}, can be implemented through Raman transitions driven by tightly focused UV addressing beams, enabling randomized benchmarking \cite{knillRandomizedBenchmarkingQuantum2008} and full programmability.

Motivated by these prospects, we envisage the following development of fermionic quantum simulators over the coming years: Short circuits of the composite gates presented here will add digital state initialization and readout capabilities to analog quantum simulators. The new observables and state preparation schemes enabled by this hybrid approach with global control will greatly enhance the utility of fermionic quantum simulators \cite{markEfficientlyMeasuring$d$wave2024, schlomerLocalControlMixed2024a, tabaresProgrammingOpticallatticeFermiHubbard2025}. With further technical progress, particularly for local control of motional quantum gates, purely digital fermionic schemes for universal computation~\cite{gonzalez-cuadraFermionicQuantumProcessing2023} or variational methods for chemistry~\cite{gkritsisSimulatingChemistryFermionic2025} will become feasible, likely based on two-dimensional superlattice architectures~\cite{zhangScalableMultipartiteEntanglement2023b}.
In the longer term, error-corrected fermionic circuits may lead to large-scale digital fermionic quantum simulation. Concrete architectures for fermionic error correction have been put forward~\mbox{\cite{schuckertFermionqubitFaulttolerantQuantum2024, ottErrorcorrectedFermionicQuantum2024}}, and we anticipate further rapid technical and conceptual developments in this active field.

~\\
\textbf{Acknowledgements:}
We thank Benjamin Schiffer, Juhi Singh, Jan A. P. Reuter, Robert Zeier for insightful discussions.
This work was supported by the Max Planck Society (MPG), the Horizon Europe program HORIZON-CL4-2022 QUANTUM-02-SGA (project 101113690, PASQuans2.1), the German Federal Ministry of Research, Technology and Space (BMFTR grant agreement 13N15890, FermiQP), and Germany's Excellence Strategy (EXC-2111-390814868). T.H. received funding from the European Research Council (ERC) under the European Union's Horizon Europe research and innovation programme (Grant Agreement No 101165353 — ERC Starting Grant FOrbQ). P.M.P acknowledges funding from the European Union’s Horizon 2020 research and innovation program (Grant agreement No 948240 — ERC Starting Grant UniRand).

~\\
\textbf{Author contributions:}
P.B. led the project, collected and analyzed most of the data.
S.W. and T.F. assisted with the data collection.
T.F. performed the Fermi-Hubbard simulations.
T.H. conceived the project.
P.B. and T.F. wrote the manuscript.
T.F., T.H. and I.B. supervised the study.
All authors worked on the interpretation of data and contributed to the final manuscript.

~\\
\textbf{Competing interests:}
The authors declare no competing interests.

~\\
\textbf{Data availability}
The datasets generated and analyzed during the present study, as well as the code used for the analysis, are available from the corresponding author upon reasonable request.

~\\
\textbf{Note:}
After preparing the manuscript, we learned from a related realization of high-fidelity quantum gates for spin exchange using fermionic atoms~\cite{KieferProtected2025}.


\bigskip

\clearpage

\clearpage
\newpage

\makeatletter

\renewcommand{\thefigure}{S\arabic{figure}}
\renewcommand{\theequation}{S\arabic{equation}}
\renewcommand{\thetable}{S\arabic{table}}
\renewcommand{\theHfigure}{S\arabic{figure}}
\renewcommand{\theHequation}{S\arabic{equation}}
\renewcommand{\theHtable}{S\arabic{table}}

\makeatother

\setcounter{figure}{0}
\setcounter{table}{0}
\setcounter{section}{0}
\setcounter{equation}{0}

\appendix

\makeatletter
\let\frontmatter@footnote@produce\relax
\let\@footnotemark\@gobble
\let\@footnotetext\@gobble
\makeatother

\title{Supplementary information for: \protect\\ High-fidelity collisional quantum gates with fermionic atoms}
\maketitle

\subsection{Experimental platform}
In our experiment, we prepare a degenerate Fermi gas of $^6\mathrm{Li}$ atoms in a balanced mixture of the two lowest hyperfine states, which represent our two spin states. The atomic cloud is loaded into a single plane of a vertical lattice following our previous work~\cite{koepsellRobustBilayerCharge2020a, hirtheMagneticallyMediatedHole2023a}, with radial confinement provided by a blue-detuned box potential projected using a digital micromirror device (DMD)~\cite{sompetRealizingSymmetryprotectedHaldane2022, hirtheMagneticallyMediatedHole2023a}. 

From there, the atoms are loaded into a 2D square optical lattice in the $x-y$ plane with lattice constants $a_{x,\,\mathrm{long}}=\SI{2.28(2)}{\micro\meter}$ and $a_{y}=\SI{1.11(1)}{\micro\meter}$. 
A DMD pattern is chosen such that a flat central region of $\sim 145$ sites is surrounded by a low-density reservoir~\cite{chalopinProbingMagneticOrigin2024}. 
The chemical potential of the reservoir, tuned by the light intensity of the DMD, controls the particle density $\langle \hat{n} \rangle$ at the center. We realize a state with an average of nearly two particles per lattice site (close to a band insulator) at lattice depths of $V_{x}^{\rm long}=9.0\,E^{\rm long}_r$ and $V_{y}=9.3\,E^{\rm short}_r$.
Dynamics are frozen by ramping the lattice depths to $V_{x}^{\rm long}=35.5\,E_r^{\rm long}$ and $V_{y}=45.0\,E_r^{\rm short}$, leaving isolated single-wells with mainly two particles per site.
Subsequently, we ramp up a second, short-spaced lattice along $x$ ($a_{x,\,\mathrm{short}}=a_{x,\,\mathrm{long}}/2$) over \SI{25}{\milli\second}, resulting in isolated, doubly occupied double-wells with total spin $S_z=0$. In this experiment, the short lattices are generated by laser beams at blue-detuned \SI{532}{\nano\meter} light incident at an angle of about $27^\mathrm{o}$. The long lattice along the $x$-direction follows the same beam path, except it is generated with red-detuned \SI{1064}{\nano\meter} light~\cite{chalopinOpticalSuperlatticeEngineering2025}.

In all figures, data points were collected in a randomized sequence to prevent systematic bias.

\subsection{State preparation fidelity}

The probability of realizing the desired state in the region of interest is approximately constant within a given dataset and primarily depends on the relative phase drift between the long and short lattices, as well as on the chosen atomic density.
It is set largely by the fidelity of preparing an average occupancy close to two atoms of opposite spin per initial lattice site, which ranges from 60\% to 85\%. 
Deviations from the target state fall into two categories: (i) empty or singly occupied double-wells, which we remove by post-selection, and (ii) double-wells containing three or more atoms, typically with population in higher lattice bands. 
Because these high-occupancy events can be mistaken for gate errors, we deliberately work at slightly lower atomic densities to suppress them, retaining between 45\% and 65\% of double-wells in analysis. 
Recent demonstrations of low-entropy band insulators in optical lattices suggest that considerably higher state preparation fidelity is attainable~\cite{chiuQuantumStateEngineering2018, yangCoolingEntanglingUltracold2020, xuNeutralatomHubbardQuantum2025}. 
We note that the state preparation step does not affect the intrinsic performance of the individual gate operations.

\begin{figure}[!b]
	\centering
	\includegraphics[scale=0.9751]{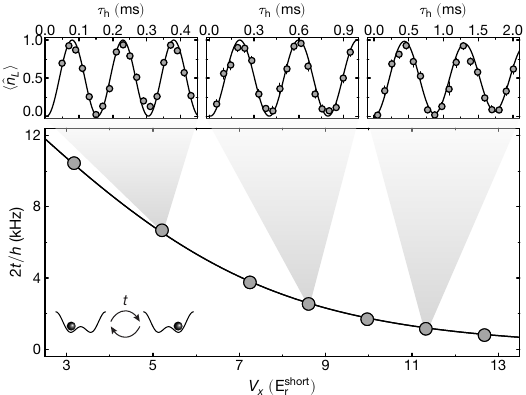}
	\caption{
		\textbf{Calibration of single particle tunneling}:
		Tunneling frequency $2t/h$ of a single atom in a double-well for different $V_{x}^{\rm short}$. The top row shows the time evolution of the population $\langle \hat{n}_L \rangle$ in the left site of the DWs as a function of holding time $\tau_{\mathrm{h}}$ following a quench to lower lattice depth at  $(\tau_\mathrm{h}=0)$. The error bars denote one s.e.m. and are smaller than the marker when not visible.
	}
	\label{fig:figS01}
\end{figure}

\subsection{Lattice depth calibration}

Lattice depth calibration is performed by measuring single-particle oscillations in a double-well, from which we extract the calibration factor by fitting the observed tunneling rates to theoretical predictions across a range of lattice depths.
An initial state consisting of a single particle in a double-well is prepared by adjusting the atom density and tilting the double-well potentials during loading, 
similar to our previous work \cite{chalopinOpticalSuperlatticeEngineering2025}.
We then remove the potential offset $\delta$, resulting in a symmetric double-well configuration at lattice depths of $V_{x}^{\rm long} = 36.5 E_r^{\rm long}$ and $(V_{x}^{\rm short}, V_y) = (56, 43)\,E_r^{\rm short}$. 
Quenching the short $x$ lattice depth to a lower value initiates coherent oscillations of the population between the two sites in the double-well. 
In our analysis, we post-select double-wells containing exactly one atom. 

In \figref{fig:figS01}, we show an example calibration plot
where the calculated calibration curve aligns with the measured tunneling frequencies with residuals below \SI {1.5}{\%} of $V_{x}^{\rm short}$.
The tunneling frequency of intra-double-well oscillations $f_t = 2t/h$ is extracted by fitting a resonant two level oscillation $\big[1 + \cos{(2\pi f_t \times \tau_{\mathrm{h}})}\big]/2$ to the population of one of the wells, which is then compared to the frequency expected from a band calculation (see our previous work~\cite{chalopinOpticalSuperlatticeEngineering2025} for more details).

\begin{figure}[!b]
	\centering
	\includegraphics[scale=0.9751]{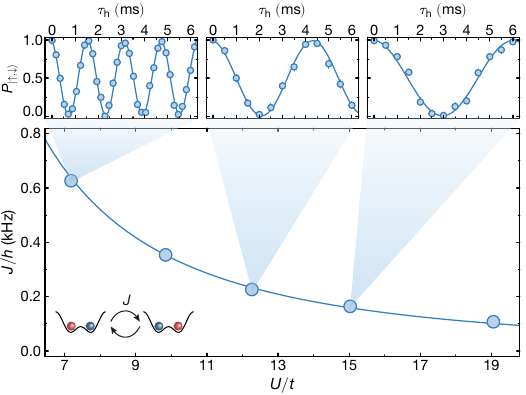}
	\caption{
		\textbf{Calibration of spin-exchange oscillations}:
		Extracted $J/h$ for different $x$-lattice depths compared to $J=4t^2/U$ (solid line). The top panels show spin-exchange oscillations at different lattice depths. 
	}
	\label{fig:figS02}
\end{figure}

To cross-check the lattice depth calibration, we measure spin-exchange oscillation in the $U/t \gg 1$ regime ($J \approx 4t^2 / U$), where virtual doublon-hole excitations are strongly suppressed (\figref{fig:figS02}). 
We compare the frequency extracted from the fit to the oscillations (\figref{fig:figS02}, upper row) with the calculated calibration curve (solid blue line) and find excellent agreement, consistent with the single-particle tunneling calibration.
The initial lattice depths in this case are $(V_{x}^{\rm short}, V_{y}) = (56, 45)\,E_r^{\rm short}$, $V_{x}^{\rm long} = 39.5\,E_r^{\rm long}$ and the Feshbach magnetic field is set to \SI{688.2}{\Gauss} to control the on-site interaction strength $U$ via a Feshbach resonance.
The long lattice depth $V_{x}^{\rm long}$ is independently calibrated using lattice-modulation spectroscopy via band-excitation energies to an accuracy level of \SI{5}{\%}.

\subsection{Experimental protocol}

The spin-exchange process is initialized from the state $\ket{\uparrow,\downarrow}$ (\figref{fig:fig3} of the main text) by linearly lowering the intra-double-well barrier from $54\,E_r^{\rm short}$ ($t\approx 0$) to $5.54\,E_r^{\rm short}$ ($t=h \times \SI{2.9(1)}{\kilo\hertz}$) in \SI{500}{\micro\s}, at on-site repulsive interactions $U=h \times \SI{6.7(1)}{\kilo\hertz}$ corresponding to a ratio $U/t\approx4/\sqrt{3}$.
After a variable holding time $\tau_\mathrm{h}$, the intra-double-well barrier is ramped back to its initial value in \SI{500}{\micro\s}.
Coherent pair-tunneling dynamics are induced under identical conditions and at the same ratio $U/t$, starting from the initial state $\ket{\uparrow\downarrow, 0}$ and with slightly modified experimental parameters (See \tabref{Tab:Exp_parameters_fig3}):

\begin{table}[h]
	\centering
	\begin{tabular}{lcc}
		\toprule
		\textbf{Parameter} & \textbf{Spin Qubit} & \textbf{DH Qubit} \\
		\midrule
		$V_x^{\rm short}$ initial & 5.54~$E_r^{\rm short}$ & 6.00~$E_r^{\rm short}$ \\
		$V_x^{\rm short}$ freezing & 54.0~$E_r^{\rm short}$ & 54.0~$E_r^{\rm short}$ \\
		$V_x^{\rm long}$ & 34.9~$E_r^{\rm long}$ & 36.8~$E_r^{\rm long}$ \\
		$V_y^{\rm short}$ & 43~$E_r^{\rm short}$ & 43~$E_r^{\rm short}$ \\
		Feshbach field & \SI{688.2}{\Gauss} & \SI{678.1}{\Gauss} \\
		\bottomrule
	\end{tabular}
	\caption{Experimental parameters of continuous spin and doublon-hole (DH) exchange oscillations (\figref{fig:fig3} of main text).}
	\label{Tab:Exp_parameters_fig3}
\end{table}

The oscillation frequency and coherence shown in \figref{fig:fig3}c and \figref{fig:fig3}e are obtained by fitting the data patches in \figref{fig:fig3}b and \figref{fig:fig3}d individually with:
\begin{equation} 
	g(\tau_\mathrm{h}) = \frac{1}{2}\Big[1 + \mathrm{A}  \cos \big(2\pi f_J(\tau_\mathrm{h} - \tau_0) \big) \Big].
\end{equation}
Here,  $\mathrm{A}$ is the contrast, $f_J = J/h$
is the  frequency of exchange oscillations and $\tau_0$ is the phase offset. 
The decay of contrast $\mathrm{A}$ (shown in \figref{fig:fig3}c and \figref{fig:fig3}e) is in both cases compatible with a Gaussian decay $\propto e^{-(\tau_\mathrm{h}/\tau_{\mathrm{ex}})^2}$ that originates from a spatial average over multiple sites with inhomogeneous oscillation frequencies ~\cite{saikoSuppressionElectronSpin2018} (see \secref{subsec:SpatialAveraging}).

\begin{table}[H]
	\centering
	\begin{tabular}{lccc}
		\toprule
		\textbf{Ramp type} & $V_{\mathrm{x}}^\mathrm{short}$ [$E_r^{\rm short}$] & $V_{\mathrm{x}}^\mathrm{long}$ [$E_r^{\rm long}$] & ROI [sites] \\ 
		\midrule
		\makecell[l]{Fast linear\\(\figref{fig:fig4}a left)} & 5.6 & 34 & 128 \\[2.5ex]
		\makecell[l]{Slow linear\\(\figref{fig:fig4}a middle)} & 5.6 & 36 & 128 \\[2.5ex]
		\makecell[l]{Blackman\\(\figref{fig:fig4}a right; 4b)} & 6.5 & 38 & 128 \\[2.5ex]
		\makecell[l]{Blackman\\(\figref{fig:fig4}c)} & 7.6 & 38 & 64 \\
	\end{tabular}
	\caption{Experimental parameters for gates with different ramp shapes shown in \figref{fig:fig4} of the main text. $V_{\mathrm{x}}^\mathrm{short}$ and $V_{\mathrm{x}}^\mathrm{long}$ are the minimal lattice depths of the short  and long lattice respectively. For all gates, the initial and final lattice depths are $V_{\mathrm{x}}^\mathrm{short} = 54\,E_r^{\rm short}$ and $V_{\mathrm{y}} = 43\,E_r^{\rm short}$; the Feshbach field is 689.9~G.}
	\label{Tab:Exp_parameters_fig4}
\end{table}

The data in \figref{fig:fig4}b and \figref{fig:fig4}c employ the quasi-adiabatic approach with Blackman pulses.
The total pulse duration for $\sqrt{\mathrm{SWAP}}$ gate is tuned to \SI{1.125}{\milli\second} in \figref{fig:fig4}b and \SI{1.29}{\milli\second} in \figref{fig:fig4}c.
The data in \figref{fig:fig4} is post-selected on having two-particles in a double-well and is not SPAM corrected.
Experimental parameters are given in \tabref{Tab:Exp_parameters_fig4}.

\subsection{Fermi-Hubbard double-well simulation}

To accurately describe the continuous exchange dynamics (\figref{fig:fig3}b and d), we simulate the Fermi-Hubbard Hamiltonian (\Eqref{eq:H_FH}) by exact diagonalization for a double-well with two particles of opposite spin with the \textit{QuSpin} library~\cite{weinbergQuSpinPythonPackage2019}. 
The calculation of the Hubbard parameters $t$ and $U$ from the depths of the optical lattices and the phase of the superlattice is explained in the Supplemental Material of~\cite{chalopinOpticalSuperlatticeEngineering2025}. 

The two-particle exchange dynamics are well reproduced by a simulation which is based on the experimental parameters in \tabref{Tab:Exp_parameters_fig3}.
Three empirical modifications are added to the bare simulation to fit the data:
First, we fine-tune the depth of the long lattice by \SI{0.3}{\%} (\SI{5}{\%}) for the spin-exchange (coherent pair-tunneling) oscillations, relative to the value expected from lattice-shaking experiments. 
Second, we observe a small chirp in the exchange frequency during the \SI{30}{\milli\second} oscillation time, which we attribute to a small gradual change of the lattice depth due to technical heating.
To account for this effect, we apply a linear correction to $V^{\rm short}$ when calculating $U$ and $t$:
\begin{equation}
	V^{\rm short}(\tau_\mathrm{h}) = V_0^{\rm short} + \Delta V^{\rm short}\  \tau_\mathrm{h}
\end{equation} 
The slope $\Delta V^{\rm short}$ was found to be $\SI{4(1)e-3}{\it{E_r^{\rm short}}\per\second}$ for both the spin exchange and the coherent pair-tunneling dynamics. 
Finally, to account for dephasing effects, the simulation results were multiplied by a Gaussian envelope, with parameters extracted from the fits shown in \figref{fig:fig3}c and \figref{fig:fig3}e. 
Aside from these three adjustments, no free parameters were needed. 
Notably, key features such as the initial phase of the oscillations and deviations from pure sinusoidal oscillations arise intrinsically from the simulation of the double-well system. 

We performed similar simulations for the different lattice ramps (\figref{fig:fig4}a). 
In this case, the only free fitting parameter is the long lattice depth $V^{\rm long}$, adjusted by \SI{0.3}{\%} in all three cases. 
Owing to the short duration of the pulses used in this experiment, thermal drifts and the associated frequency chirp can be safely neglected and were therefore not included in the simulation. 

To reproduce the Ramsey oscillations of \figref{fig:fig5}c in simulation, we increase the experimental long lattice depth by \SI{12}{\%} to calibrate the $U_{\rm int}(\pi/2)$ pulses. Because the idealized simulation does not capture all residual inhomogeneities, we also reduce the simulated contrast by \SI{8.7}{\%}, a value extracted from a sinusoidal fit to the data.
The experimental parameters used for the pair-exchange gates in \figref{fig:fig5}e are identical to those of \figref{fig:fig5}c, except for the short-lattice depth during the $3\pi/2$ pulse. This depth is reduced to $3.3\,E^{\rm short}_{r}$, optimized so that the $\ket{\uparrow,\downarrow}$ and $\ket{\downarrow, \uparrow}$ initial states undergo the desired $3\pi/2$ rotation. 
With only this modification compared to \figref{fig:fig5}c, the resulting simulation reproduces the experimental data of \figref{fig:fig5}e with good agreement.
This comparison confirms that the reduced contrast is primarily caused by direct-exchange processes, which become prominent at the very low lattice depths used there and slightly modify the effective exchange coupling $J$ in the spin and charge sectors.
At higher lattice depths, where direct exchange is negligible, these effects are suppressed, and uniformly high performance for all initial states is achievable.

\begin{figure}[!t]
	\centering
	\includegraphics[scale=0.9751]{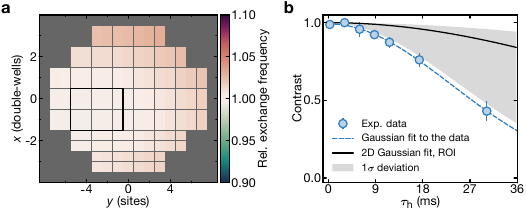}
	\caption{
		\textbf{Ensemble dephasing due to spatial frequency inhomogeneity}:
		\textbf{a,} Spatial map of the relative spin-exchange frequency, obtained via a 2D Gaussian fit to the data shown in \figref{fig:fig3}c. The black rectangle marks the  region of interest (ROI) taken into account for the oscillation plots in \figref{fig:fig3}b and the simulation.
		\textbf{b,} Contrast decay as a function of hold time $\tau_\mathrm{h}$. 
		Blue circles show experimental data, with a Gaussian fit (dashed blue line) indicating coherence decay. The solid black line shows the simulated contrast decay for the 2D Gaussian obtained from the fit, while the gray error band shows the range of results when shifting this fit by the 68\% fitting uncertainties in the $(x_0,y_0)$ center position.
	}
	\label{fig:figS03}
\end{figure}

\subsection{Effect of spatial averaging on collisional gates}
\label{subsec:SpatialAveraging}

The decay of the global spin-exchange contrast (\figref{fig:fig3}c,e) arises from inhomogeneous local oscillation frequencies, which lead to a Gaussian envelope upon averaging over multiple sites~\cite{saikoSuppressionElectronSpin2018}.
This behavior is further supported by comparing the experimental data to simulations that incorporate site-resolved distributions of spin-exchange frequencies. To capture the spatial inhomogeneity, the relative spin-exchange frequency map from \figref{fig:fig3}c is fitted with a two-dimensional (2D) Gaussian profile (\figref{fig:figS03}a). 
Averaging over this fitted spatial distribution yields the contrast decay shown by the black curve shown in \figref{fig:figS03}b. 
The gray error band represents the range of simulated outcomes obtained by shifting the center position ($x_0$, $y_0$) of the 2D Gaussian fit within its 68\% confidence interval. 
This result reproduces both the Gaussian form and the correct order of magnitude of the decay of contrast, confirming its consistency with inhomogeneous dephasing. 
Additional sources of dephasing such as lattice disorder or temporal fluctuations are not included in the model and may further reduce contrast.

\subsection{Two-qubit fidelity estimate}
The fidelity $F_{\sqrt{\mathrm{SWAP}}}$ of the entangling gate is estimated from an exponential decay fit $P(N_p) = p_0\left(F_{\sqrt{\mathrm{SWAP}}}\right)^{{N_p}}$ (\figref{fig:fig4}c), where $p_0$ is the initial state population, and ${N_p}$ is the number of applied pulses.

\begin{figure}[!t]
	\centering
	 \includegraphics[scale=0.9751]{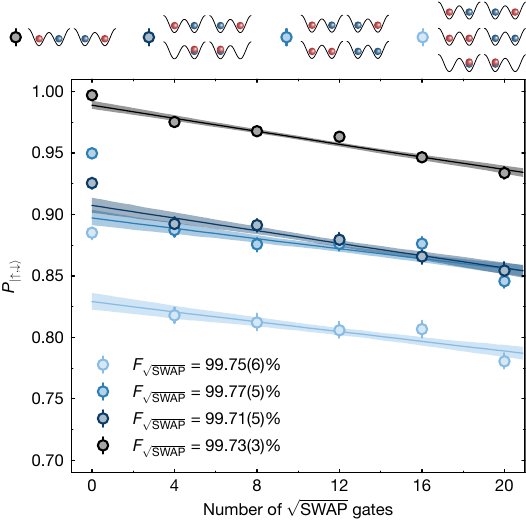}
	\caption{\textbf{Two-qubit gate fidelity}: 
		We extract fidelities for different states remaining after post-selection, as shown by the cartoons in the top. The solid lines of the respective color show the exponential fits.
	}
	\label{fig:figS04}
\end{figure}

With our fully spin and charge-resolved imaging, the two-qubit gates errors depend on states kept in the analysis i.e. the chosen qubit basis. 
In a pure spin quantum computer, all measured states involving doublons or holes can be trivially ignored, while in a full fermionic quantum computer all states are physically relevant and contribute to the error of the gate.  
\figref{fig:figS04} shows how the fidelity estimation depends on this choice.
In the most general case for two-particle states, we post-select on having two particles in one double-well potential (\figref{fig:figS04} light blue circles and \figref{fig:fig4}c). 
For a spin-qubit basis, the unphysical states are $\ket{\uparrow\downarrow,0}$ and $\ket{0,\uparrow\downarrow}$, while on the other hand states $\ket{\uparrow, \uparrow}$ and $\ket{\downarrow, \downarrow}$ are not part of the $S_z=0$ basis. 
Black circles correspond to post-selection of only $\ket{\uparrow,\downarrow}$ and $\ket{\downarrow, \uparrow}$ states.
Extracted fidelities are shown in the legend and largely remain unaffected by post-selection.

\begin{figure}[!t]
	\centering
	 \includegraphics[scale=0.9751]{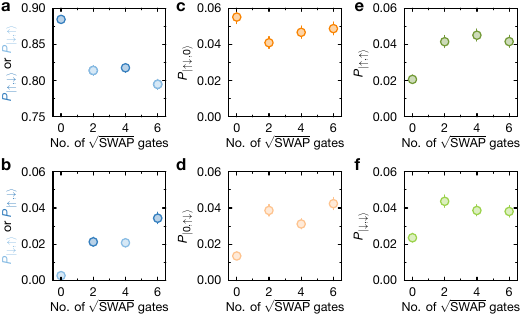}
	\caption{\textbf{State preparation errors}: Measured fraction of two-particle states in \textbf{a} / \textbf{b,} $\ket{\uparrow,\downarrow}$ or $\ket{\downarrow, \uparrow}$, \textbf{c,} $\ket{\uparrow\downarrow,0}$ \textbf{d,} $\ket{0, \uparrow\downarrow}$ \textbf{e,} $\ket{\uparrow,\uparrow}$ \textbf{f,} $\ket{\downarrow,\downarrow}$, as a function of number of $\sqrt{\mathrm{SWAP}}$ pulses, for the initial state $\ket{\uparrow,\downarrow}$. Two $\sqrt{\mathrm{SWAP}}$ gates perform a $\mathrm{SWAP}$ operation leading to alternating entries in \textbf{a, b}. Some state-preparation errors like band-excitations are only detected after some time resulting initial drop in \textbf{a}. } 
	\label{fig:figS05}
\end{figure}

The jump in the $\ket{\uparrow,\downarrow}$ population after applying the first entangling pulse can be explained by a state preparation error that is not captured by the post-selection. 
During the initial preparation, which should lead to two particles with opposite spins per site, it can happen for two atoms with identical spin states to occupy the same lattice site, residing in different motional bands. 
Following spin-dependent splitting used for initial-state preparation, such configurations (e.g., $\ket{\uparrow, \uparrow}$ or $\ket{\downarrow, \downarrow}$) are distributed in the excited band (one well) and the ground band (the other well).
Upon vertical spin-splitting used for the final detection, these atoms are displaced in opposite directions due to their band-dependent motion, making them indistinguishable from the target state $\ket{\uparrow,\downarrow}$.
If a gate pulse is applied before vertical spin splitting, the atoms in higher bands tunnel out of the double-well
and throughout the system, and are wrongly detected as one of the unwanted two-particle state, which is removed by post-selection.  
This is clearly visible in \figref{fig:figS05}, which shows fractions of particles in each of the six two-particle states for the initial state $\ket{\uparrow, \downarrow}$: The population in the initial state drops after applying the first pulse, while numbers in almost all other states increase at this step. The data point without any gates is thus omitted in the determination of the gate fidelity.

\begin{figure}[!t]
	\centering
	 \includegraphics[scale=0.9751]{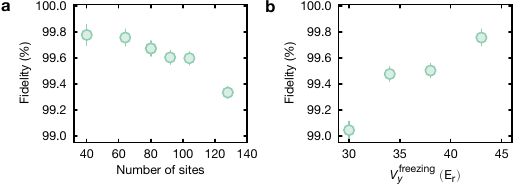}
	\caption{\textbf{Finite-size and lattice-depth effects on fidelity}: 
		\textbf{a,} Dependence of mean fidelity on number of considered lattice sites. 
		\textbf{b,} Mean fidelity of 64 lattice sites for different freezing lattice depths, showing that we are not yet in the saturation regime.
	}
	\label{fig:figS06}
\end{figure}

As shown in \figref{fig:figS03}, the measured fidelity is limited by the system’s homogeneity and thus depends on the system size. Scaling the 64-qubit system to 128 lattice sites results in a slight decrease in average fidelity to 99.3\% (\figref{fig:figS06}a). In future experiments, larger system sizes and higher fidelities could be achieved by employing larger lattice beams and flattening the potential using a digital micromirror device (DMD)~\cite{sompetRealizingSymmetryprotectedHaldane2022, hirtheMagneticallyMediatedHole2023a}.

Throughout this work, the limited maximum $y$-lattice depth has been among the leading sources of gate infidelity because residual inter-well tunneling can lead to gate errors or misidentification of the final states.
In \figref{fig:figS06}b, we show the dependence of mean fidelity of the central 64 lattice sites for different maximal lattice depths. 
We find that the freezing lattice depth of $43\,E_r^{\rm short}$ is still on the rising slope of fidelity.
Increasing the lattice depth provides a direct route to further improve fidelities and reduce particle losses.

\begin{figure}[!b]
	\centering
	\includegraphics[scale=0.9751]{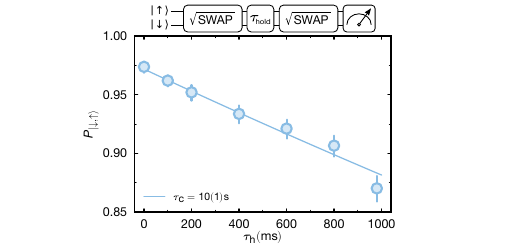}
	\caption{
		\textbf{Coherence of Bell state}:
		Phase stability test of the entangled state using a variable hold time between two $\sqrt{\mathrm{SWAP}}$ pulses. The first pulse prepares the state from $\ket{\uparrow,\downarrow}$ and the second maps accumulated phase into population for readout.
	}
	\label{fig:figS07}
\end{figure}

\subsection{Dephasing protection of spin qubits}

The dephasing protection of spin qubits originates from their low sensitivity to magnetic field gradients. 
At a Feshbach field of \SI{688.0}{\Gauss}, the energetically lowest two $\mathrm{^6Li}$ spin states exhibit a differential magnetic moment of $\Delta\mu_{\uparrow-\downarrow} \approx \SI{5}{\kilo \hertz \per \Gauss}$. 
For dephasing to occur, an energy difference between the product states $\ket{\uparrow,\downarrow}$ and $\ket{\downarrow,\uparrow}$ is needed, which scales as $\Delta E \propto \Delta\mu_{\uparrow\downarrow }\cdot\Delta B$, where $\Delta B$ is a magnetic field gradient.

One way to test for unwanted phase evolution is shown in \figref{fig:figS07}. 
After preparing the Bell state \mbox{$(\ket{\uparrow,\downarrow}+i\ket{\downarrow,\uparrow})/\sqrt{2}$}, we freeze the dynamics for variable hold times and then apply a disentangling pulse that maps the atoms onto $P_{\ket{\downarrow,\uparrow}}$.
Fitting an exponential decay yields a decoherence timescale of \SI{10(1)}{\second}. 
This timescale is limited by dephasing due to residual magnetic field gradients, hence it serves as a lower bound on the actual coherence of the Bell state. 
This lower bound on the coherence of the system exceeds the \SI{1.3}{\milli\second} required for a single entangling pulse by four orders of magnitude, meaning that spin-qubit decoherence has insignificant contributions to collisional gate fidelity.

In a second Ramsey experiment, we measure coherence via singlet-triplet oscillations in a magnetic field gradient, as shown in \figref{fig:fig4}d. 
We observe oscillations of the population at a frequency of \SI{8.72(5)}{\hertz} that are compatible with the expected $\Delta E=h \times \Delta\mu_{\uparrow - \downarrow} \Delta B$ for a magnetic gradient of $\Delta B = \SI{40.1(1)}{\Gauss\cm^{-1}}$. 
During the measurement time of \SI{1}{\second}, we observe negligible decay of the oscillation contrast. To quantify coherence time, we fit the data with both exponential and Gaussian decay models, yielding decoherence times of \SI{125}{\second} (\SI{28}{\second}), with 68\%-confidence intervals from \SI{25}{\second} (\SI{5}{\second}) to infinity. We use the profile likelihood method from the \textit{lmfit} library~\cite{newvilleLMFITNonLinearLeastSquares2025} to estimate these confidence intervals. 
Based on these measurements, we can conclude a conservative lower bound of \SI{10}{\second} on the coherence of the spin Bell state.

\subsection{Sequence design and control parameters for interaction and pair-exchange gate \texorpdfstring{$PX(\Theta)$}{PX(Theta)}}

\begin{figure}[H]
	\centering
	 \includegraphics[scale=0.9751]{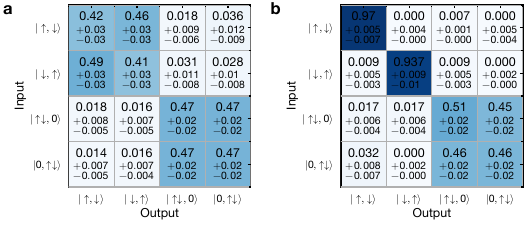}
	\caption{
		\textbf{Truth table for four input states}:
		\textbf{a,} for interaction gate, without SPAM error correction.
		\textbf{b,} for pair-exchange gate ($PX(\Theta)$), without SPAM error correction.
	}
	\label{fig:figS08}
\end{figure}

The interaction gate $U_{\mathrm{int}}(\pi/2)$ in \figref{fig:fig5}a is realized by lowering the $V_{x}^{\rm short}$ from $54.0\,E_r^{\rm short}$ to $7.87\,E_r^{\rm short}$ in \SI{0.6}{\milli\second}, with $V_{x}^{\rm long} = 35.0\,E_r^{\rm long}$. 
The lattice depth ramp is shaped as a quadratic pulse, which, similar to the Blackman pulse, helps mitigate doublon excitations and offers a robust, experimentally convenient pulse shape.
The analysis is limited to three double-wells where we post-select on having two particles in one double-well, the truth table is not SPAM corrected. Since part of the data is obscured in \figref{fig:fig5}a, we also present the data in \figref{fig:figS08}a.

\begin{figure}[!t]
	\centering
	 \includegraphics[scale=0.9751]{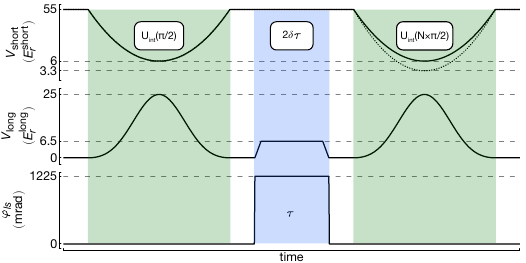}
	\caption{
		\textbf{Time traces of pulses}:
		Experimental protocol for a charge-sensitive Ramsey (solid line, $N=1$) and $\mathrm{PX}(\Theta)$ gate sequence (dotted line, $N=3$) that is robust to fluctuations and gradients of the phase ($\varphi_{\mathrm{ls}}$) between long and short lattices. For the interaction gate (green rectangles), we ramp down the short lattice depth $V^{\rm short}$ to induce inter-well tunneling, and ramp up the long lattice depth $V^{\rm long}$ to confine the atoms in the double-wells. 
		The total duration of $U_{\rm int}$ gates is \SI{1}{\milli \s}.
		For the $Z$-gate tilt (blue rectangle), we use a shallow long lattice and large relative phase $\varphi_{\mathrm{ls}}$ between long and short lattice to realize a controlled energy difference $\delta$ for varying duration $\tau$.
	}
	\label{fig:figS09}
\end{figure}

For composite pulse sequences such as the Ramsey sequence shown in \figref{fig:fig5}c and the pair-exchange gate illustrated in \figref{fig:fig5}e and f, precise control of the relative phase $\theta$ between the states $\ket{\uparrow\downarrow, 0}$ and $\ket{0, \uparrow\downarrow}$ is essential. 
This relative phase is directly linked to the bias $\delta$, which scales as $\delta \propto V_{x}^{\rm long} \sin(\varphi_{\mathrm{ls}})$. In a standard approach, where the long lattice depth is held constant throughout the gate sequence, fluctuations or spatial gradients in the relative phase $\varphi_{\mathrm{ls}}$ between the long and short lattice potentials limit our performance. To mitigate this, we design an improved pulse sequence, shown in \figref{fig:figS09}, that is more robust against such unwanted fluctuations:
To avoid that the phase $\theta$ accumulates outside of the gate time, the long lattice is here off, as $\delta$ scales with the long lattice depth $V_{x}^{\rm long}$. 
Using a similar protocol to the one in \figref{fig:fig5}c, for the interaction gate (green rectangle), we ramp down the short lattice depth to induce intra-double-well tunneling, and ramp up the long lattice depth to confine the atoms in the double-wells. 
For the charge sensitive $Z$-gate tilt (blue rectangle), we employ a shallow long lattice with a large lattice phase $\varphi_{\mathrm{ls}}$. Since the error in $\varphi_{\mathrm{ls}}$ is absolute, using a large phase suppresses the error. The optimal choice is $\varphi_{\mathrm{ls}} = \pi/2$, where the sensitivity to fluctuations is only quadratic. However, due to technical constraints, the experiment was conducted at $\varphi_{\mathrm{ls}}=0.3\,\pi$.

\subsection{Outlook and prospects for the experimental platform}

\begin{figure}[!t]
	\centering
	 \includegraphics[scale=0.9751]{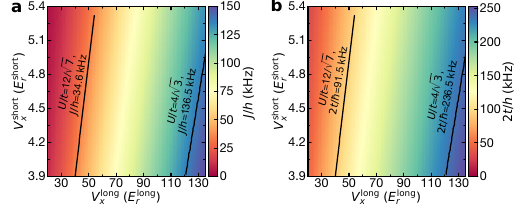}
	\caption{
		\textbf{Expected parameters in an improved platform}:
		\textbf{a,} Expected spin-exchange frequency $J$ and \textbf{b,} Tunneling frequency $t$ for a lattice spacing of \SI{383.5}{\nano\meter}, lattice depths of $V_y=45\,E_r^{\rm short}$ and $V_z=50\,E_r^{\rm z}$
		and Feshbach field of \SI{767.2}{\Gauss}. 
		Black lines represent magic $U/t$ ratios.}
	\label{fig:figS10}
\end{figure}

The optical superlattice platform offers substantial scope for further advancement of system size \cite{yangCoolingEntanglingUltracold2020} and gate performance discussed in this work.
In terms of scalability and gate speed, combining light fermionic $^6\mathrm{Li}$ with three times shorter lattice spacings of \SI{383.5}{\nano\meter} (already demonstrated in a quantum-gas microscope~\cite{impertroUnsupervisedDeepLearning2023}), significantly faster quantum gates and usable array sizes on the order of $10^4$ lattice sites become realistic. 
Band-structure calculations and a generalized spin-exchange expression (see \secref{subsec:AnalyticDerivation}) give conservative estimates of \SI{135}{\kilo\hertz} for the spin-exchange rate and \SI{235}{\kilo\hertz} for single-particle tunneling (see \figref{fig:figS10}), indicating that sub-\SI{10}{\micro\second} gates are feasible. 
Optimal-control pulse shaping could further shorten these times \cite{nemirovskyFastUniversalTwoqubitGate2021, singhOptimizingTwoqubitGates2025}, while randomized benchmarking~\cite{knillRandomizedBenchmarkingQuantum2008} will provide a comprehensive assessment of gate fidelity.

\subsection{Composition of the pair-exchange gate}

\figref{fig:fig5}e illustrates the composite pair-exchange (PX) gate implemented in this work, which consists of a phase gate $U_Z(\theta)$ sandwiched between two interaction gates.
The $Z$-phase pulse is applied via the bias $\delta$ of the double-well (see Hamiltonian in \Eqref{eq:H_FH} of the main text) and results in:
\begin{align}
	U_{\mathrm{Z}}(\Theta) \nonumber
	= \left[
	\begin{array}{cccccc}
			1 & 0 & 0 & 0 & \multicolumn{2}{c}{\multirow{4}{*}{\raisebox{-0em}{\large$\mathbf{0}$}}} \\
			0 & 1 & 0 & 0 & \multicolumn{2}{c}{} \\
			0 & 0 & 1 & 0 & \multicolumn{2}{c}{} \\
			0 & 0 & 0 & 1 & \multicolumn{2}{c}{} \\
			\multicolumn{4}{c}{\multirow{2}{*}{\raisebox{0em}{\large$\mathbf{0}$}}} & e^{-i\Theta - i\zeta''} & 0\\
			\multicolumn{4}{c}{} & 0 & e^{i\Theta - i\zeta''}
		\end{array}
	\right]\,.
\end{align} 
Here, $\Theta = 2\pi\times \delta \tau_\mathrm{hz}/h$ is the tilt phase arising from the energy offset $\delta$, and $\zeta'' = 2\pi\times U \tau_\mathrm{hz}/h$ is dependent on the on-site interaction.

The composite sequence \textit{Int}–\textit{Z}–\textit{Int}, comprising two interaction gates and one \(Z\)-phase pulse, results in:
\begin{align}
	\mathrm{PX}(\Theta) &= U_{\mathrm{int}}\left(\tfrac{3\pi}{2}\right) \cdot U_{\mathrm{Z}}(\Theta) \cdot U_{\mathrm{int}}\left(\tfrac{\pi}{2}\right) \nonumber \\
	&= \left[
	\begin{array}{cccccc}
			1 & 0 & 0 & 0 & \multicolumn{2}{c}{\multirow{4}{*}{\raisebox{-0em}{\large$\mathbf{0}$}}} \\
			0 & 1 & 0 & 0 & \multicolumn{2}{c}{} \\
			0 & 0 & 1 & 0 & \multicolumn{2}{c}{} \\
			0 & 0 & 0 & 1 & \multicolumn{2}{c}{} \\
			\multicolumn{4}{c}{\multirow{2}{*}{\raisebox{0em}{\large$\mathbf{0}$}}} & e^{-i\zeta'}\cos(\Theta) & -e^{-i\zeta'}\sin(\Theta) \\
			\multicolumn{4}{c}{} & e^{-i\zeta'}\sin(\Theta) & e^{-i\zeta'}\cos(\Theta)
		\end{array}
	\right]\,.
\end{align}
Here $\zeta'=2\pi\times U \tau_\mathrm{total}/h$  is a $U$-dependent phase associated with the combined duration $\tau_\mathrm{total}$ of the three applied gates. This phase can be effectively canceled by appending an appropriate waiting time at the end of the sequence. 
Because the tilt \(\delta\) couples exclusively to the doublon-hole (DH) manifold and is invisible to the spin manifold, the protocol isolates pair-exchange from background spin-exchange.

\begin{figure}[!t]
	\centering
		\includegraphics[scale=0.9751]{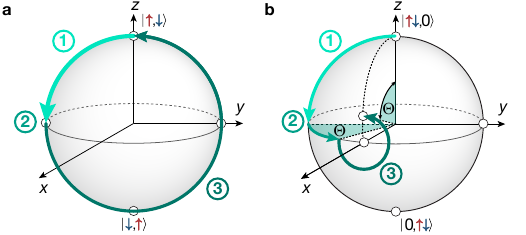}
	\caption{\textbf{Illustration of the composite pulse for the pair-exchange gate ($PX(\Theta)$)}: Pulses 1 and 3 are interaction gates $U_\mathrm{int}(\theta_i)$ with $\theta_1=\pi/2$ and $\theta_3=3\pi/2$. Pulse 2 is a $Z$-gate $U_Z(\Theta)$.
		\textbf{a,} In the spin sector, pulse 2 has no effect on the states. Thus, pulses 1 and 3 together implement  a full $2\pi$ rotation around the Bloch sphere. 
		\textbf{b,} In the charge sector, pulse 1 rotates the initial state $\ket{\uparrow\downarrow, 0}$ by $\pi/2$ around the $x$-axis of the Bloch sphere. 
		In a general case, pulse 2 then rotates the state along the equator by an angle $\Theta$. 
		The final pulse 3 acts as a rotation around the $x$-axis.
		For $\Theta=(2N+1)\pi/2$ this sequence realizes a pair-exchange gate.
	}
	\label{fig:figS11}
\end{figure}

\figref{fig:figS11} shows the corresponding Bloch-sphere trajectories: the doublon-hole manifold (\figref{fig:figS11}b) traces a great-circle arc of angle \(\Theta\), whereas the spin manifold (\figref{fig:figS11}a) executes a closed loop and returns to its origin.  
These trajectories verify that the composite sequence realizes the intended pair-exchange operation with high fidelity while leaving the spin sector untouched. 
\figref{fig:figS08}b shows the truth table for the diagram in \figref{fig:fig5}f for $\mathrm{PX}(\Theta=7\pi/2)$.

\subsection{Fermi-Hubbard double-well analytical derivation}
\label{subsec:AnalyticDerivation}

The Fermi-Hubbard Hamiltonian 
\begin{equation}
	\hat{H}_{\mathrm{FH}} = -t\sum_{\sigma\in\{\uparrow,\downarrow\}} \left(\hat{c}_{\mathrm{L}, \sigma}^{\dagger} \hat{c}_{\mathrm{R}, \sigma} + \mathrm{h.c.}\right) + U \sum_{i\in\{\mathrm{L}, \mathrm{R}\}} \hat{n}_{i, \uparrow}\hat{n}_{i, \downarrow}
	\label{eq:H_FH_sup}
\end{equation}
for a double-well with two particles of opposite spin can be expressed in matrix form in the basis 
\begin{align}
	\left\{
	\begin{array}{l}
		\ket{\psi_1} = \ket{\uparrow, \downarrow} = \hat{c}^\dagger_{L,\uparrow}\hat{c}^\dagger_{R,\downarrow}\ket{0} \\[4pt]
		\ket{\psi_2} = \ket{\downarrow, \uparrow} = \hat{c}^\dagger_{L,\downarrow}\hat{c}^\dagger_{R,\uparrow}\ket{0} = -\hat{c}^\dagger_{R,\uparrow} \hat{c}^\dagger_{L,\downarrow}\ket{0}\\[4pt]
		\ket{\psi_3} = \ket{\uparrow\downarrow, 0} = \hat{c}^\dagger_{L,\uparrow}\hat{c}^\dagger_{L,\downarrow}\ket{0}\\[4pt]
		\ket{\psi_4} = \ket{0, \uparrow\downarrow} = \hat{c}^\dagger_{R,\uparrow}\hat{c}^\dagger_{R,\downarrow}\ket{0}
	\end{array}
	\right.
\end{align}
as
\begin{equation}
	\hat{H}_{\mathrm{FH}}^{\mathrm{dw}, S_z=0} = \left[\begin{smallmatrix}
		0 & 0 & -t & -t \\
		0 & 0 & t & t \\
		-t & t & U & 0 \\
		-t & t & 0 & U \\
	\end{smallmatrix}\right].
\end{equation}
This matrix can be diagonalized, yielding the eigenvalues
\begin{equation}
	\lambda_1 = 0, \quad
	\lambda_2 = U, \quad
	\lambda_3 = -J, \quad
	\lambda_4 = U + J
\end{equation}
and eigenvectors
\begin{align*}
	\mathbf{e}_1 &= \frac{1}{\sqrt{2}} \left[\begin{smallmatrix}
		1 \\
		1 \\
		0 \\
		0
	\end{smallmatrix}\right], \qquad \qquad
	\mathbf{e}_2 = \frac{1}{\sqrt{2}}\left[\begin{smallmatrix}
		0 \\
		0 \\
		-1 \\
		1
	\end{smallmatrix}\right], \\
	\mathbf{e}_3 &= \dfrac{1}{2\sqrt{\Ueff}}\left[\begin{smallmatrix}
		\sqrt{\Ueff + U} \\
		-\sqrt{\Ueff + U} \\
		\sqrt{\Ueff - U} \\
		\sqrt{\Ueff - U}
	\end{smallmatrix}\right]
	\overset{t/U \rightarrow 0}{\rightarrow} \frac{1}{\sqrt{2}} \left[\begin{smallmatrix}
		1 \\
		-1 \\
		0 \\
		0
	\end{smallmatrix}\right],\\
	\mathbf{e}_4 &= \dfrac{1}{2\sqrt{\Ueff}}\left[\begin{smallmatrix}
		-\sqrt{\Ueff - U} \\
		\sqrt{\Ueff - U} \\
		\sqrt{\Ueff + U} \\
		\sqrt{\Ueff + U}
	\end{smallmatrix}\right]\overset{t/U \rightarrow 0}{\rightarrow} \frac{1}{\sqrt{2}} \left[\begin{smallmatrix}
		0 \\
		0 \\
		1 \\
		1
	\end{smallmatrix}\right],
\end{align*}
where we have introduced the effective interaction 
\begin{equation}
	\Ueff = \sqrt{U^2 + 16t^2}
	\label{eq:Ueff}
\end{equation} 
and generalized spin-exchange 
\begin{equation}
	J=\nicefrac{1}{2}(\Ueff - U).
	\label{eq:J}
\end{equation} 
Expanding this in small $\epsilon=t/U$ gives $\Ueff = U + 2\tilde{J} + \mathcal{O}(\epsilon^4)$ and $J = \tilde{J} + \mathcal{O}(\epsilon^4)$ with $\tilde{J} = 4t^2/U$. 

\begin{figure}[!t]
	\includegraphics[scale=0.9751]{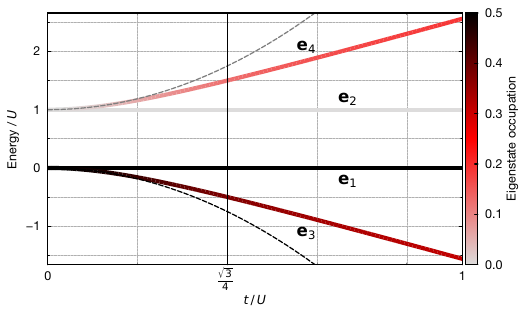}
	\caption{
		\textbf{Eigenvalues of the Fermi-Hubbard Hamiltonian}:
		Eigenvalues $\lambda_i$ as a function of $t/U$ for the four eigenvectors $\mathbf{e}_i$ in a double-well with two particles and total spin projection $S_z=0$. The colorbar denotes the eigenstate occupation of $|\braket{\uparrow, \downarrow}{\mathbf{e_i}}|^2$ for all four eigenstates $\ket{\mathbf{e_i}}$. The dashed lines represent the approximations to $\lambda_3$ and $\lambda_4$ in the limit $t/U\ll 1$. Marked as a vertical line is the magic ratio $t/U = \sqrt{3}/4$.
	} 
	\label{fig:figS12}
\end{figure}

It helps intuition to analyze the eigensystem (see \figref{fig:figS12}) in the limit of small hopping-to-interaction ratio (\( t/U \ll 1 \)), where the eigenstates naturally separate into spin- and charge-dominated sectors. 
The lowest two eigenstates, \( \mathbf{e}_1 \) and \( \mathbf{e}_3 \), are predominantly spin-like (eigenstate occupations $|\braket{\uparrow, \downarrow}{\mathbf{e_i}}|^2 \approx |\braket{\downarrow, \uparrow}{\mathbf{e_i}}|^2 \approx \frac{1}{2}$, see also colorbar in \figref{fig:figS12}) and are adiabatically connected as \( t\rightarrow 0 \). The state \( \mathbf{e}_1 \) corresponds to a spin-triplet configuration without any doublon occupation, while \( \mathbf{e}_3 \) exhibits only a small admixture of doubly occupied states. 
Consequently, the spin dynamics are primarily governed by the coherent evolution between these two states and occur at a characteristic energy scale \( J \), given by their energy splitting. 
In contrast, the states \( \mathbf{e}_2 \) and \( \mathbf{e}_4 \) are charge-dominated (eigenstate occupations $|\braket{\uparrow, \downarrow}{\mathbf{e_i}}|^2 \approx |\braket{\downarrow, \uparrow}{\mathbf{e_i}}|^2 \approx 0$, see also colorbar in \figref{fig:figS12}), composed mainly of configurations with one site doubly occupied and the other empty (\( \ket{\uparrow\downarrow, 0} \) and \( \ket{0, \uparrow\downarrow} \)). 
These charge excitations also exhibit an energy separation on the order of \( J \), but are energetically offset from the spin sector by approximately \( U \), the on-site interaction energy. 
This separation defines a clear hierarchy of energy scales, allowing the spin and charge dynamics to be treated independently at leading order. 

A natural protocol for initiating coherent dynamics is to prepare the system in a state \( \ket{\psi_m}\) and perform a sudden quench of the tunneling amplitude from zero to a finite value $t > 0$. 
Following the quench, each eigenstate $\mathbf{e}_i$ acquires a dynamical phase, resulting in the time evolved state (with $\hbar = h/(2\pi)$)
\begin{equation}
	\matrixel{\psi_m}{e^{-i \hat{H}_{\mathrm{FH}}\tau_\mathrm{h}/\hbar}}{\psi_n} = \sum_i \braket{\psi_m}{\mathbf{e}_i}\braket{\mathbf{e}_i}{\psi_n}e^{-i\lambda_i \tau_\mathrm{h}/\hbar }.
	\label{eq:general_time_evolution}
\end{equation}
It is instructive to visualize this time evolution as a sum of rotating vectors in the complex plane. Each vector corresponds to an eigenstate $\mathbf{e}_i$, with its amplitude given by the product of wavefunction overlaps $\left| \braket{\psi_m}{\mathbf{e}_i} \braket{\mathbf{e}_i}{\psi_n} \right|$, and its phase evolving at frequency $\lambda_i$. 

\begin{figure}[t!]
	\includegraphics[scale=0.9751]{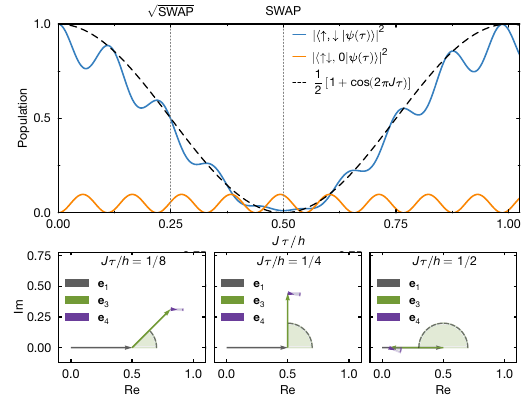}
	\centering
	\caption{
		\textbf{Time evolution of $\ket{\psi_0} = \ket{\uparrow, \downarrow}$ for $U/t= 5$ and for a quench of the lattice depth}: 
		Top: Time evolution of the population in $\ket{\uparrow, \downarrow}$ (blue) and in the doublon state $\ket{\uparrow\downarrow, 0}$ (orange). 
		The dashed black line shows an oscillation at frequency $J$. 
		Lower panels: Visualization of the time evolution as a sum of rotating vectors in the complex plane. 
		Each vector represents an eigenstate $\mathbf{e}_i$, with an amplitude given by the product of wavefunction overlaps $|\langle \psi_m | \mathbf{e}_i \rangle \langle \mathbf{e}_i | \psi_n \rangle|$ and a phase that evolves at the eigenfrequency $\lambda_i$. 
	} 
	\label{fig:figS13}
\end{figure}

As a concrete example, consider the matrix element $\matrixel{\uparrow,\downarrow}{e^{-i \hat{H}_{\mathrm{FH}} \tau_\mathrm{h}/\hbar}}{\uparrow,\downarrow}$. This evolution is determined by three vector contributions (see \figref{fig:figS13}). The first, associated with eigenstate $\mathbf{e}_1$, has an overlap amplitude of $\frac{1}{2}$ and an eigenvalue $\lambda_1 = 0$, meaning this vector remains stationary over time. For $t/U \ll 1$, the second most significant contribution arises from eigenstate $\mathbf{e}_3$, rotating at frequency $J$. This term drives the spin-exchange dynamics. The overlap of $\ket{\uparrow,\downarrow}$ with $\mathbf{e}_2$ is exactly zero, and with $\mathbf{e}_4$ is zero in the limit of $t/U\ll1$. In marysummary, evaluating \Eqref{eq:general_time_evolution} in this limit, leads to 
\begin{equation}
	\begin{aligned}
		&\hspace{-1em}
		\matrixel{\uparrow,\downarrow}{e^{-i \hat{H}_{\mathrm{FH}} \tau_\mathrm{h}/\hbar}}{\uparrow, \downarrow} \\[3pt]
		&\mathrel{\overset{\smash{t/U\ll1}}{\approx}}\; \left|\braket{\uparrow,\downarrow}{\mathbf{e}_1}\right|^2 e^{-i \lambda_1 \tau_\mathrm{h}/\hbar}
		+ \left|\braket{\uparrow,\downarrow}{\mathbf{e}_3} \right|^2
		e^{-i \lambda_3 \tau_\mathrm{h}/\hbar} \\[3pt]
		&\mathrel{\overset{\smash{t/U\ll1}}{\approx}}\;  \frac{1}{2} + \frac{1}{2} e^{i J \tau_\mathrm{h}/\hbar}.
	\end{aligned}
	\label{eq:population_swap}
\end{equation}
Similarly, the complete time evolution operator can be derived as:
\begin{widetext}
	\begin{equation}
		U_{\mathrm{int}}(\theta=J\tau_\mathrm{h}/\hbar) \overset{t/U\ll1}=
		\left[\begin{smallmatrix}
			\frac{1 + e^{iJ\tau_\mathrm{h}/\hbar}}{2} & \frac{1 - e^{iJ\tau_\mathrm{h}/\hbar}}{2} & 0 & 0\\
			\frac{1 - e^{iJ\tau_\mathrm{h}/\hbar}}{2} & \frac{1 + e^{iJ\tau_\mathrm{h}/\hbar}}{2} & 0 & 0\\
			0 & 0 & \mathrm{e}^{-i U \tau_\mathrm{h}/\hbar}\frac{1 + e^{-iJ\tau_\mathrm{h}/\hbar}}{2} & -\mathrm{e}^{-i U \tau_\mathrm{h}/\hbar}\frac{1 - e^{-iJ\tau_\mathrm{h}/\hbar}}{2}\\
			0 & 0 & -\mathrm{e}^{-i U \tau_\mathrm{h}/\hbar}\frac{1 - e^{-iJ\tau_\mathrm{h}/\hbar}}{2} & \mathrm{e}^{-i U \tau_\mathrm{h}/\hbar}\frac{1 + e^{-iJ\tau_\mathrm{h}/\hbar}}{2}
		\end{smallmatrix}\right],
		\label{eq:XXZZgatefull}
	\end{equation}
\end{widetext}
Which is exactly the lower 4x4 block of the interaction matrix $U_{\mathrm{int}}(\theta)$ from equation (2) with $\theta=J\tau_\mathrm{h}/\hbar$ and $\zeta=U\tau_\mathrm{h}/\hbar$.
Note that the Fermi-Hubbard Hamiltonian (\Eqref{eq:H_FH_sup}) preserves the projection of the total spin $\hat{S}_z$, therefore it does not couple the states $\ket{\uparrow,\uparrow}$ and $\ket{\downarrow,\downarrow}$ to any other state. The exchange process described in our work couples two states that do not directly interact via a virtual intermediate state that is energetically detuned. This results in a second-order, effectively coherent two-body interaction.\\

Errors to the $\mathrm{SWAP}^\alpha$ gate arise firstly from a small contribution from eigenstate $\mathbf{e}_4$ whose vector rotates rapidly with frequency $U + J$ (see gray vector in the bottom row in \figref{fig:figS13}). This high-frequency component introduces a phase and amplitude error that scales as $\Ueff - U\approx 2\tilde{J} + U \mathcal{O}(\epsilon^4)$. In addition, for the initial state $\ket{\uparrow, \downarrow}$, the doublon population $|\braket{\uparrow\downarrow, 0}{\uparrow, \downarrow}|^2$ oscillates coherently with frequency $\Ueff$ and peak-to-peak amplitude $4t^2/\Ueff^2$ (see orange line in \figref{fig:figS13}):
\begin{equation}
	\begin{aligned}
		&\hspace{-2em}
		\matrixel{\uparrow\downarrow, 0}{e^{-i \hat{H}_{\mathrm{FH}} \tau_\mathrm{h}/\hbar}}{\uparrow, \downarrow} \\[3pt]
		&=\; \braket{\uparrow\downarrow, 0}{\mathbf{e}_3} \braket{\mathbf{e}_3}{\uparrow, \downarrow} e^{-i \lambda_3 \tau_\mathrm{h}/\hbar} \\[4pt]
		&\quad \quad +\; \braket{\uparrow\downarrow, 0}{\mathbf{e}_4} \braket{\mathbf{e}_4}{\uparrow, \downarrow} e^{-i \lambda_4 \tau_\mathrm{h}/\hbar} \\[4pt]
		&=\; \frac{t}{U_\mathrm{eff}} e^{-\frac{i}{2}(U - U_\mathrm{eff})\tau_\mathrm{h}/\hbar}
		- \frac{t}{U_\mathrm{eff}} e^{-\frac{i}{2}(U + U_\mathrm{eff})\tau_\mathrm{h}/\hbar} \\[4pt]
		&=\; \frac{2i\, t}{U_\mathrm{eff}}\,
		e^{-\frac{iU\tau_\mathrm{h}}{2\hbar}}\,
		\sin\!\left( \frac{U_\mathrm{eff} \tau_\mathrm{h}}{2\hbar} \right)
	\end{aligned}
	\label{eq:doublon_population}
\end{equation}
This equation holds generally for arbitrary $U/t$ because $\Ueff=\lambda_4-\lambda_3$ is the energy difference of the two populated eigenstates with doublon-hole population. 

To speed up the gates, the ratio $U/t$ can be reduced, however this increases the influence of the fast-rotating $\mathbf{e}_4$ vector, leading to a substantial phase and amplitude error of the $\mathrm{SWAP}^\alpha$ gate unless $\Ueff\cdot\tau_\mathrm{h}$ is an integer multiple of $2\pi$~\cite{yangCoolingEntanglingUltracold2020}. 
This leads to a series of \textit{magic ratios} for $U/t$. For an entangling $\pi/2$ pulse, the gate time is $\tau_\mathrm{h}=\hbar\pi/(2J)$ leading to the condition 
\begin{align*}
	\Ueff\cdot\tau_\mathrm{h}/\hbar&=\frac{\Ueff\,\pi}{2J}\stackrel{!}{=}2\pi n\\
	\Rightarrow\left(\frac{U}{t}\right)_\textrm{magic $\pi/2$}&=\frac{4(2n-1)}{\sqrt{4n-1}},
\end{align*}
where the integer $n$ gives the number of full $2\pi$ rotations of the
doublon-hole population during the time of the $\pi/2$ exchange gate.
Explicitly, $\Rightarrow\left(\frac{U}{t}\right)_\textrm{magic $\pi/2$} \in \{4/\sqrt{3}, \; 12/\sqrt{7}, \; 20/\sqrt{11}...\}$, enable the realization of maximally entangled Bell states. For arbitrary $\alpha=\theta/\pi$ the \textit{magic ratio} pulse is generalized to:
\begin{equation*}
	\left(\frac{U}{t}\right)_{\pi \alpha}=\frac{4(n-\alpha)}{\sqrt{\alpha(2n-\alpha)}},
\end{equation*}

\begin{figure}[t!]
	\includegraphics[scale=0.9751]{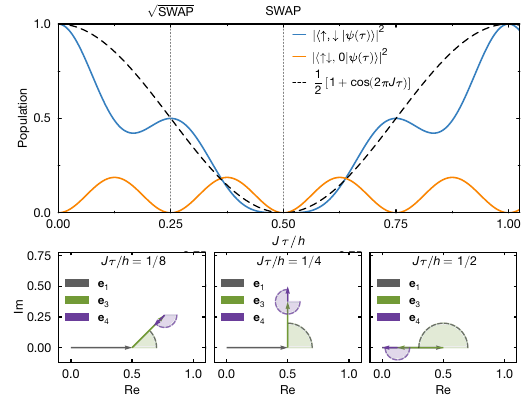}
	\centering
	\caption{\textbf{Time evolution of $\ket{\psi_0} = \ket{\uparrow, \downarrow}$ for the magic ratio $U/t=4/\sqrt{3}$}: Otherwise same as  \figref{fig:figS13}.}
	\label{fig:figS14}
\end{figure}

The magic ratio with the fastest dynamics, and thus the fastest gate operation of this protocol, is $U/t=4/\sqrt{3}$, for which the effective interaction is $U_\mathrm{eff} = 2U = 8t/\sqrt{3}$, and spin-exchange is $J = U/2 = 2t/\sqrt{3}$. This quantifies how the collisional gate speed is ultimately set by the interaction strength.
Evaluating all matrix elements in \Eqref{eq:general_time_evolution} for $U/t=4/\sqrt{3}$ and $\tau_\mathrm{h} = h/(4J)$ again yields the interaction matrix 
\[
U_\mathrm{int} \overset{\substack{\frac{U}{t}=\frac{4}{\sqrt{3}} \\ \tau_\mathrm{h}=\frac{h}{4J}}}{=} \begin{pmatrix}
	\frac{1}{2} + \frac{i}{2} & \frac{1}{2} - \frac{i}{2} & 0 & 0 \\[6pt]
	\frac{1}{2} - \frac{i}{2} & \frac{1}{2} + \frac{i}{2} & 0 & 0 \\[6pt]
	0 & 0 & -\frac{1}{2} + \frac{i}{2} & \frac{1}{2} + \frac{i}{2} \\[6pt]
	0 & 0 & \frac{1}{2} + \frac{i}{2} & -\frac{1}{2} + \frac{i}{2}
\end{pmatrix}
\] which is again exactly the lower 4x4 diagonal block of $U_{\mathrm{int}}(\theta)$ with $\theta=J\tau_\mathrm{h}/\hbar = \frac{\pi}{2}$ and $\zeta = U\tau_\mathrm{h}/\hbar = \pi$.  
The same analysis as in the previous $U/t \gg 1$ is shown in \figref{fig:figS14}.
In theory, this condition provides an optimal realization of the $\sqrt{\mathrm{SWAP}}$ gate. However, in practice, it is experimentally demanding to fine-tune both the interaction ratio $U/t$ and the pulse duration with the required precision.

\begin{figure}[!b]
	\includegraphics[scale=0.9751]{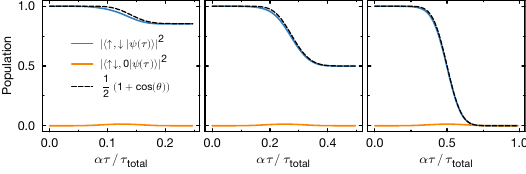}
	\centering
	\caption{\textbf{Blackman Pulses}: Time evolution of the population in $\ket{\uparrow, \downarrow}$ (blue solid line) and $\ket{\uparrow\downarrow, 0}$ (orange solid line) for quasi-adiabatic pulses with $\alpha \in \{\frac{1}{4}, \frac{1}{2}, 1\}$ starting in $\ket{\uparrow, \downarrow}$. For all pulses, the tunneling amplitude $t$ is ramped according to a Blackman profile, while the on-site interaction $U$ remains constant, with ratio of $U/t > 8$. For all three resulting gate angles, the final state has negligible doublon population, and the population in $\ket{\uparrow, \downarrow}$ is well described by~\Eqref{eq:theta_pulse} (black dashed line). }
	\label{fig:figS15}
\end{figure}

An alternative approach to implementing the gate is to ramp the tunneling amplitude $t$ quasi-adiabatically from zero to its final value. Provided the ramp is slow compared to the interaction scale $U$, the system evolves within the subspace spanned by the eigenstates $\mathbf{e}_1$ and $\mathbf{e}_3$, which remain predominantly populated. Consequently, the amplitude of the vector associated to $\mathbf{e}_4$ remains zero throughout the evolution. This eliminates amplitude and phase errors, as well as any unwanted coupling between the spin and charge sectors.
Importantly, there is no requirement for a minimal gate speed: since $\mathbf{e}_1$ and $\mathbf{e}_3$ are degenerate at $t = 0$ the overlap with these two states is always exactly equal. Additionally, the specific shape of the pulse is inconsequential, provided it remains sufficiently slow. The resulting gate angle $\alpha$ is determined solely by the total duration of the pulse $\tau_{\mathrm{total}}$ and the time-dependent exchange interaction $J(\tau_\mathrm{p})$, according to:
\begin{equation}
	\alpha = \frac{1}{\pi}\int_0^{\tau_{\mathrm{total}}} J(\tau_\mathrm{p})/\hbar d\tau_\mathrm{p}.
	\label{eq:theta_pulse}
\end{equation}
\figref{fig:figS15} shows the evolution of the populations during Blackman pulses of different length demonstrating the suppression of doublon-hole population.

\end{document}